\documentclass[twocolumn,astrosymb,tighten]{aastex63}
\usepackage{comment}
\usepackage{CJK}
\usepackage{savesym}
\savesymbol{tablenum}
\usepackage{siunitx}
\restoresymbol{SIX}{tablenum}
\DeclareSIUnit\year{yr}
\DeclareSIUnit\parsec{pc}
\DeclareSIUnit\erg{erg}

\newcommand{\pwater}{H$_3$O$^+$}
\newcommand{\gal}{NGC\,253}
\submitjournal{The Astrophysical Journal}

\shorttitle{Cosmic Ray Ionization Rate in NGC\,253}
\shortauthors{Holdship et al.}
\graphicspath{{./}{./figures/}}

\begin{document}
\begin{CJK*}{UTF8}{gbsn}

\title{Energizing Star Formation: The Cosmic Ray Ionization Rate in NGC\,253 Derived From ALCHEMI Measurements of H$_3$O$^+$ and SO}

\author[0000-0003-4025-1552]{Jonathan Holdship}
\affiliation{Leiden Observatory, Leiden University, PO Box 9513, NL-2300 RA Leiden, The Netherlands}
\affiliation{Department of Physics and Astronomy, University College London, Gower Street, London WC1E 6BT}
\author[0000-0003-1183-9293]{Jeffrey G.~Mangum}
\affiliation{National Radio Astronomy Observatory, 520 Edgemont Road,
  Charlottesville, VA  22903-2475, USA}
%
\author[0000-0001-8504-8844]{Serena Viti}
\affiliation{Leiden Observatory, Leiden University, PO Box 9513, NL-2300 RA Leiden, The Netherlands}
\affiliation{Department of Physics and Astronomy, University College London, Gower Street, London WC1E 6BT}
\author[0000-0002-2333-5474]{Erica Behrens}
\affiliation{Department of Astronomy, University of Virginia, P.~O.~Box 400325, 530 McCormick Road, Charlottesville, VA 22904-4325}
\author[0000-0002-6824-6627]{Nanase Harada}
\affiliation{National Astronomical Observatory of Japan, 2-21-1 Osawa, Mitaka, Tokyo 181-8588, Japan}
\affiliation{Department of Astronomy, School of Science, The Graduate University for Advanced Studies (SOKENDAI), 2-21-1 Osawa, Mitaka, Tokyo, 181-1855 Japan}

\author[0000-0001-9281-2919]{Sergio Mart\'in}
\affiliation{European Southern Observatory, Alonso de C\'ordova, 3107, Vitacura, Santiago 763-0355, Chile}
\affiliation{Joint ALMA Observatory, Alonso de C\'ordova, 3107, Vitacura, Santiago 763-0355, Chile}

\author[0000-0001-5187-2288]{Kazushi Sakamoto}
\affiliation{Institute of Astronomy and Astrophysics, Academia Sinica, 11F of AS/NTU
Astronomy-Mathematics Building, No.1, Sec. 4, Roosevelt Rd, Taipei 10617, Taiwan}

\author[0000-0002-9931-1313]{Sebastien Muller}
\affiliation{Department of Space, Earth and Environment, Chalmers University of Technology, Onsala Space Observatory, SE-43992 Onsala, Sweden}

\author[0000-0001-8153-1986]{Kunihiko Tanaka}
\affil{Department of Physics, Faculty of Science and Technology, Keio University, 3-14-1 Hiyoshi, Yokohama, Kanagawa 223--8522 Japan}

\author[0000-0002-6939-0372]{Kouichiro Nakanishi}
\affiliation{National Astronomical Observatory of Japan, 2-21-1 Osawa, Mitaka, Tokyo 181-8588, Japan}
\affiliation{Department of Astronomy, School of Science, The Graduate University for Advanced Studies (SOKENDAI), 2-21-1 Osawa, Mitaka, Tokyo, 181-1855 Japan}

\author[0000-0002-7758-8717]{Rub\'en~Herrero-Illana}
\affiliation{European Southern Observatory, Alonso de C\'ordova, 3107, Vitacura, Santiago 763-0355, Chile}
\affiliation{Institute of Space Sciences (ICE, CSIC), Campus UAB, Carrer de Magrans, E-08193 Barcelona, Spain}

\author{Yuki Yoshimura}
\affiliation{Institute of Astronomy, Graduate School of Science,
The University of Tokyo, 2-21-1 Osawa, Mitaka, Tokyo 181-0015, Japan}




\author[0000-0002-1316-1343]{Rebeca Aladro}
\affiliation{Max-Planck-Institut f\"ur Radioastronomie, Auf dem H\"ugel 69, 53121 Bonn, Germany}

 \author[0000-0001-8064-6394]{Laura Colzi}
\affiliation{Centro de Astrobiolog\'ia (CSIC-INTA), Ctra. de Ajalvir Km. 4, 28850, Torrej\'on de Ardoz, Madrid, Spain}
\affiliation{INAF-Osservatorio Astrofisico di Arcetri, Largo E. Fermi 5, I-50125, Florence, Italy 5}





\author[0000-0001-6527-6954]{Kimberly L. Emig}
\affiliation{National Radio Astronomy Observatory, 520 Edgemont Road,
  Charlottesville, VA  22903-2475, USA}

 \author[0000-0002-7495-4005]{Christian Henkel}
 \affiliation{Max-Planck-Institut f\"ur Radioastronomie, Auf dem H\"ugel   69, 53121 Bonn, Germany}
 \affiliation{Astronomy Department, Faculty of Science, King Abdulaziz
   University, P.~O.~Box 80203, Jeddah 21589, Saudi Arabia}







 \author[0000-0003-0563-067X]{Yuri Nishimura}
 \affiliation{Institute of Astronomy, The University of Tokyo, 
 2-21-1, Osawa, Mitaka, Tokyo 181-0015, Japan}
 \affiliation{ALMA Project, National Astronomical Observatory of Japan, 
 2-21-1, Osawa, Mitaka, Tokyo 181-8588, Japan}

 \author[0000-0002-2887-5859]{V\'ictor M.~Rivilla}
 \affiliation{Centro de Astrobiolog\'ia (CSIC-INTA), Ctra. de Ajalvir Km. 4, 28850, Torrej\'on de Ardoz, Madrid, Spain}
 \affiliation{INAF-Osservatorio Astrofisico di Arcetri, Largo E. Fermi 5, I-50125, Florence, Italy 5}



 \author[0000-0001-5434-5942]{Paul P.~van der Werf}
 \affiliation{Leiden Observatory, Leiden University,
     PO Box 9513, NL - 2300 RA Leiden, The Netherlands}
 

\collaboration{20}{(ALMA Comprehensive High-resolution Extragalactic Molecular Inventory (ALCHEMI) collaboration)} 

\correspondingauthor{J. Holdship, holdship@strw.leidenuniv.nl}



\begin{abstract}

The cosmic ray ionization rate (CRIR) is a key parameter in understanding the physical and chemical processes in the interstellar medium. Cosmic rays are a significant source of energy in star formation regions, which impacts the physical and chemical processes which drive the formation of stars.  Previous studies of the circum-molecular zone (CMZ) of the starburst galaxy \gal{} have found evidence for a high CRIR value; $10^3-10^6$ times the average cosmic ray ionization rate within the Milky Way.  This is a broad constraint and one goal of this study is to determine this value with much higher precision.
We exploit ALMA observations towards the central molecular zone of \gal{} to measure the CRIR. We first demonstrate that the abundance ratio of \pwater and SO is strongly sensitive to the CRIR . We then combine chemical and radiative transfer models with nested sampling to infer the gas properties and CRIR of several star-forming regions in \gal{} from emission from their transitions.
We find that each of the four regions modelled has a CRIR in the range $(1-80)\times10^{-14}$\,s$^{-1}$ 
and that this result adequately fits the abundances of other species that are believed to be sensitive to cosmic rays including C$_2$H, HCO$^+$, HOC$^+$, and CO. From shock and PDR/XDR models, we further find that neither UV/X-ray driven nor shock dominated chemistry are a viable single alternative as none of these processes can adequately fit the abundances of all of these species.
\end{abstract}

\keywords{starburst galaxies,Galaxies: ISM, Galaxies: active, Galaxies: abundances}


\section{Introduction} 
\label{sec:intro}
Cosmic rays play an important role in the interstellar medium as a source of heating and ionization. They drive chemistry by ionizing atoms; many gas-phase reaction chains begin with ionization followed by a barrierless reaction \citep{williams2013}. Moreover, interactions with cosmic rays heat the gas \citep{Goldsmith2001}, which can spur more complex chemistry and also affects the dynamics of the system in question.\par
It is clear that measuring the cosmic ray ionization rate (CRIR) is vital to understanding the chemical and dynamical evolution of molecular clouds.  Careful measurement of the CRIR is required in order to properly characterize the energy budget within star formation regions. For this reason, there have been many efforts to measure the CRIR in the Milky Way \citep[e.g.,][]{Padovani2009,Indriolo2015} and in extragalactic environments \citep[e.g.,][]{Gonzalez-Alfonso2013,Muller2016}. These measurements typically use ratios of OH$^+$, H$_2$O$^+$, and H$_3$O$^+$ to infer the CRIR assuming chemical equilibrium has been reached.\par
In this work, we attempt to infer the average CRIR for several regions in the central molecular zone (CMZ) of the starburst galaxy \gal{} using other molecular line ratios. Due to its rich molecular emission, \gal{} was selected as the target of the ALMA Comprehensive High-Resolution Extragalactic Molecular Inventory (ALCHEMI), an ALMA large program \citep{Martin2021}. The goal of ALCHEMI is to produce the most complete molecular inventory of an extragalactic object and use that to drive understanding of the CMZ.\par
The CMZ of \gal{} contains several large ($\sim 50$\,pc), dense ($\sim$\SI{e5}{\per\centi\metre\cubed}), well-studied molecular clouds \citep[e.g.,][]{Sakamoto2011,Leroy2018} which are labelled in Fig.~\ref{fig:H3Op320221IntSO3423Int}. Often labelled as giant molecular clouds (GMCs) due to their size, they have much higher masses and have higher velocity dispersions \citep{Leroy2015} than typical GMCs in the Milky Way. We refer to these GMC-like structures as GMCs for simplicity throughout this work.  See \cite{Leroy2015} for position information associated with these GMCs, and \cite{Mangum2019} for the association of these GMCs with other identified dense gas positions in NGC\,253. It is for the brightest of these GMCs (3, 4, 5, 6 and 7), closest to the centre of the CMZ, that we intend to infer the CRIR.\par
Previous work has attempted to measure the CRIR in the GMCs of \gal{}. For example, \citet{Holdship2021} found that a high C$_2$H abundance in the GMCs was likely caused by cosmic rays. However, an attempt to use chemical modelling to measure the CRIR found that it could only be constrained in the range \SIrange[]{e-14}{e-11}{\per\second} due to the low sensitivity of the C$_2$H abundance to the CRIR. Meanwhile, \citet{Harada2021} found that HCO$^+$ and HOC$^+$ measurements indicated a cosmic ray ionization rate  \textgreater \SI{e-14}{\per\second} confirming - but not further constraining - the previous measurement. Despite these large uncertainties, it is clear the CRIR in these regions is much higher than the Galactic value which is typically measured to be in the range \SIrange{1e-17}{10e-17}{\per\second} \citep{Padovani2009,Indriolo2015}. More sensitive tracers are required to obtain robust and accurate measurements of the CRIR in star forming regions. In this work, we consider \pwater{} and SO since their abundance ratio has previously been found to be strongly dependent on the CRIR \citep{Bayet2011}. We first confirm this dependency using a large grid of chemical models and then use emission from these species to infer the CRIR in the CMZ of \gal{}.\par
In Sect.~\ref{sec:data}, the data reduction steps followed to extract line intensities from the ALCHEMI image cubes are described. In Sect.~\ref{sec:approach} we present preliminary modelling to justify our analysis and the approach used to infer the CRIR for the targeted GMCs of \gal{}. In Sect.~\ref{sec:results} we present the results of the analysis which is discussed in Sect.~\ref{sec:discussion}.  We present our summary in Sect.~\ref{sec:conclusion}.
\section{Observational Data}
\label{sec:data}
\subsection{ALCHEMI Data}
\label{sec:alchemidata}
We make use of data acquired as part of the ALCHEMI ALMA large program. A full description of the ALCHEMI observations and data reduction can be found in \citet{Martin2021} but important details are given here. ALCHEMI is an unbiased spectral survey of the central molecular zone (CMZ) of the starburst galaxy \gal{} covering ALMA bands 3 through 7 (84 to 373\,GHz). \par
\begin{figure*}[htbp]
\centering
\includegraphics[trim=0mm 0mm 0mm 0mm, clip, scale=0.64]{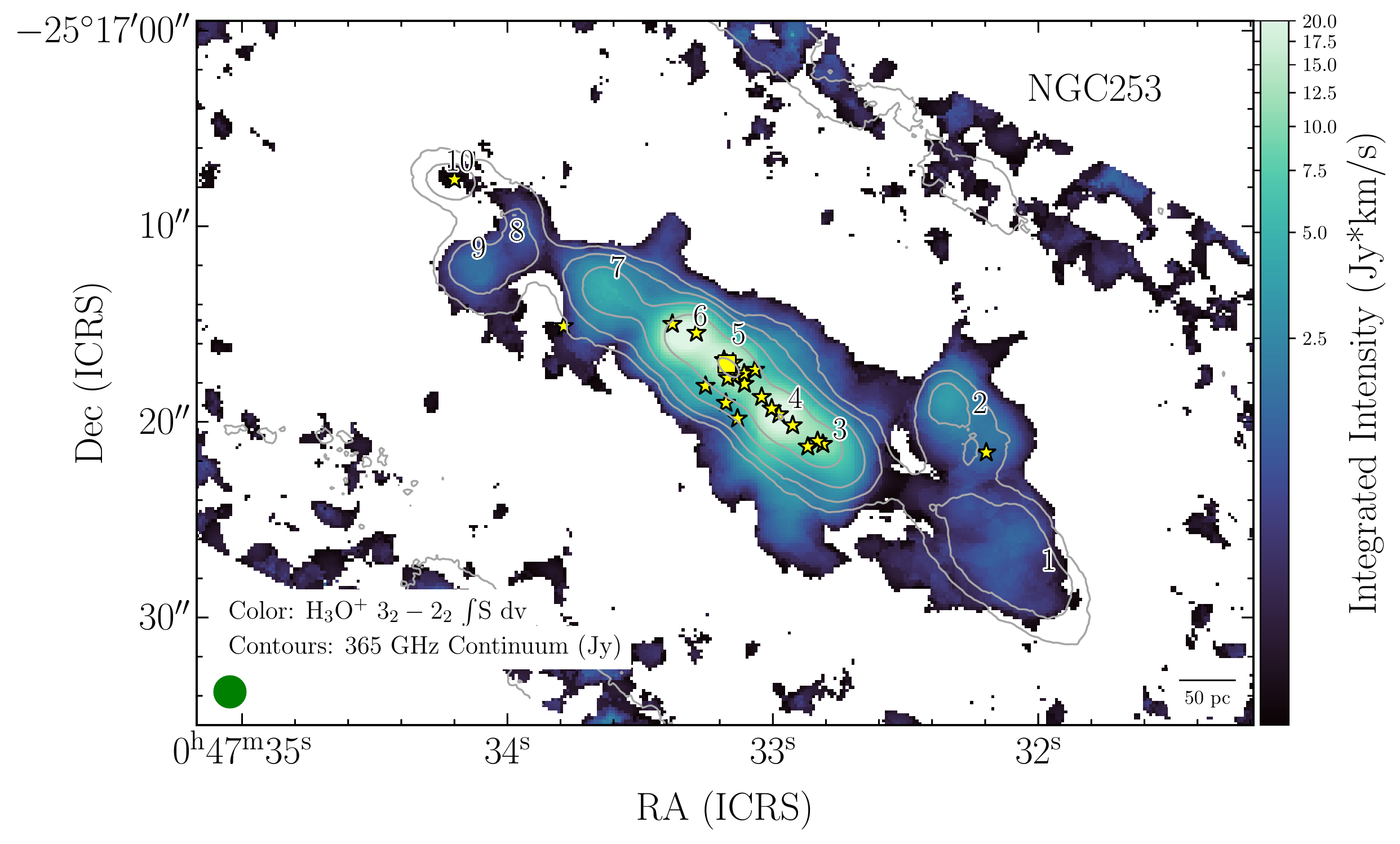}\\
\includegraphics[trim=0mm 0mm 0mm 0mm, clip, scale=0.64]{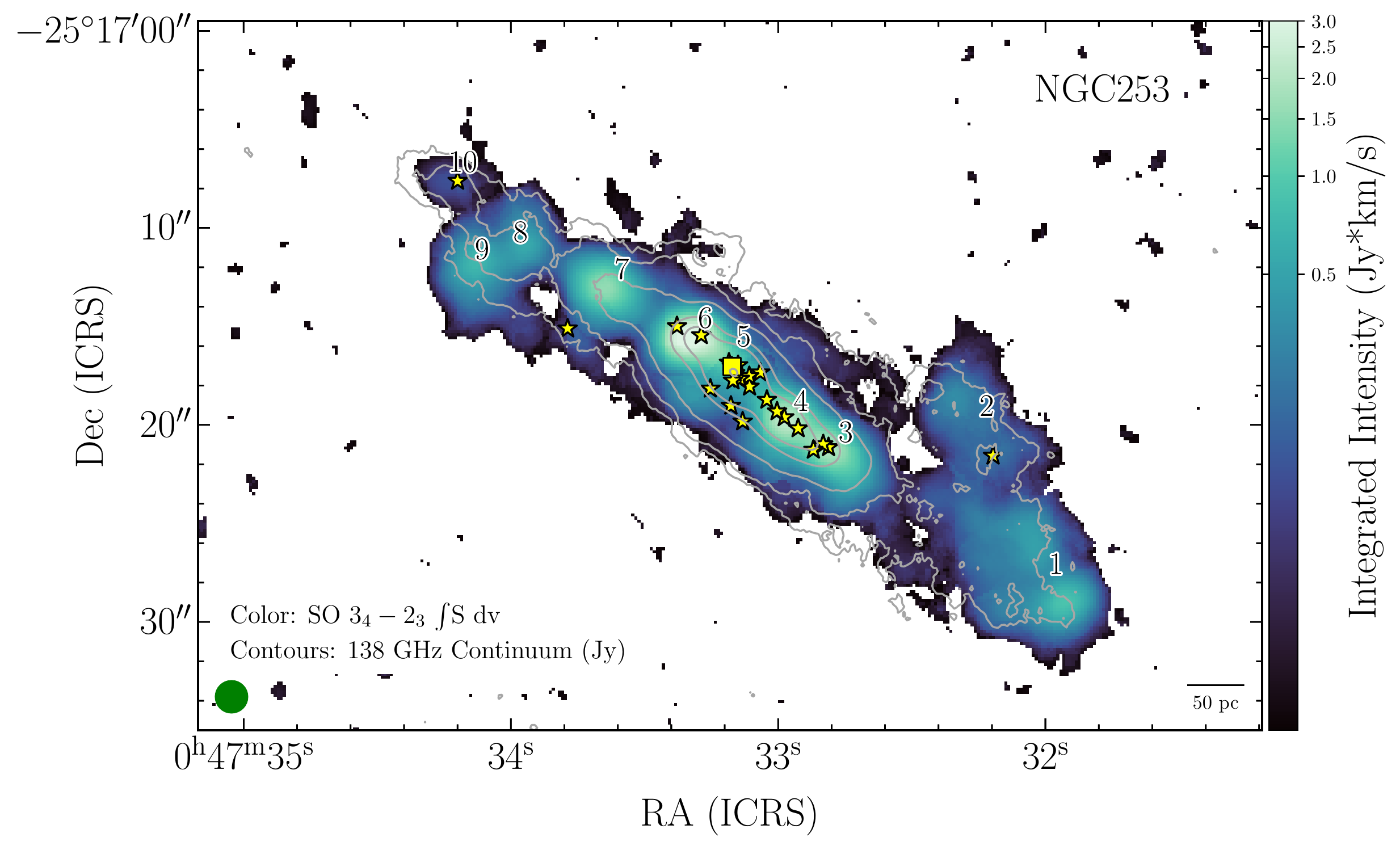}
\caption{Sample \pwater{} (top) and SO (bottom) integrated intensity (moment 0) images toward NGC\,253.  For each image the green ellipse and black scale bar in the lower-left and lower-right corners show the final imaged beam size (1.6\,arcsec) and physical scale, respectively.  Red numbers indicate the locations of the dense molecular emission regions identified by \citet[][Table 4]{Leroy2015}. Star shaped markers locate the positions of the 2\,cm radio continuum emission peaks \citep{Ulvestad1997}, with a square indicating the position of the strongest radio continuum peak identified by \citet{Turner1985}. The lower integrated intensity limit for each transition is set to 3$\sigma$ in the integrated intensity (see Section~\ref{momentanal}). Overlain in contours is the associated continuum emission distribution for each transition.  Continuum contours are in steps of 3, 6, 9, 12, 30, 120, 240, and 900 times the respective continuum RMS, where the peak continuum intensity dictates the number of these levels actually used for a given panel.  The continuum RMS values for the \pwater{} and SO transitions shown are 1.0 and 0.15\,mJy/beam, respectively.}
\label{fig:H3Op320221IntSO3423Int}
\end{figure*}
\gal{} was observed toward a nominal phase center of $\alpha=00^h47^m33.26^s$,  $\delta= -25^{\circ}17'17.7''$ (ICRS). Observations were configured to cover a common rectangular area of $50\arcsec\times20\arcsec$ with a position angle of $65^\circ$ (East of North). All ALCHEMI image cubes presented in this article were imaged to a common beam size of 1\farcs6.  The maximum recoverable scale for the ALCHEMI observations is 15\arcsec. If the distance of \gal{} is taken to be 3.5~Mpc \citep{Rekola2005}, these angular size scales correspond to linear size scales of 28 and 250\,pc respectively. The spectral range of the ALCHEMI data includes 37 SO and 2 \pwater{} rotational transitions.  These SO and \pwater{} transitions are listed with their frequencies and quantum numbers in Table~\ref{table:transitions}.  The recommended $1\sigma$ absolute flux calibration uncertainty for measurements from the ALCHEMI survey is 15\% \citep{Martin2021}, a value which we adopt in the analysis presented in this article.  We have also extracted the continuum emission associated with the ALCHEMI measurements listed in Table~\ref{table:transitions} using the continuum subtraction and imaging process described in \citet{Martin2021}.\par
\startlongtable
\begin{deluxetable*}{lclcl}
\tablewidth{0pt}
\tablecaption{SO\tablenotemark{a} and \pwater\ Transitions and Frequencies\label{table:transitions}}
\tablehead{
\colhead{Target} &
\colhead{Frequency / GHz} &
\colhead{E$_{\mathrm{U}}$ / K} &
\colhead{Interloper} &
\colhead{Overlap Correction\tablenotemark{b}}
}
\decimalcolnumbers
\startdata
H$_3$O$^+$ $1_1-2_1$ & 307.192 & 79.5 & CH$_3$OH $4_1-4_0$ & (0.25,0.23,0.21,0.21,0.24) \\
H$_3$O$^+$ $3_2-2_2$ & 364.797 & 139.8& OCS $30-29$ & (1.0,0.86,0.88,0.83,1.0) \\
 & . & & HC$_3$N v$_7$=1 40-39 & (1.0,1.0,0.71,1.0,1.0) \\
H$_3$O$^+$ $3_0-2_0$ & 396.272 & 169.1& & 1.0 \\
SO $2_2-1_1$ & 86.093 & 19.3 & HC$^{15}$N $1-0$ & 1.0 \\
SO $2_3-1_2$ & 99.299 & 9.22 & NH$_2$CN $5_15-4_14$ & 1.0 \\
SO $5_4-4_4$ & 100.029 & 38.6 & HC$_3$N $11-10$ & 1.0 \\
SO $3_2-2_1$ & 109.252 & 21.1 & HC$_3$N $12-11$ & 1.0 \\
SO $3_3-2_2$ & 129.138 & 25.5 & & 1.0 \\
SO $6_5-5_5$ & 136.634 & 50.7 & & 1.0 \\
SO $3_4-2_3$ & 138.178 & 15.9 & & 1.0 \\
SO $4_3-3_2$ & 158.971 & 28.7 & & 1.0 \\
SO $4_4-3_3$ & 172.181 & 33.7 & HC$^{15}$N $2-1$ & 1.0 \\
SO $4_5-3_4$ & 178.605 & 24.4 & & 1.0 \\
SO $5_4-4_3$ & 206.176 & 38.6 & CCS $16_15-15_14$ & 1.0 \\
SO $8_7-7_7$ & 214.357 & 81.2 & H$_2^{34}$S $2_20-2_11$ & 0.0 \\
SO $5_5-4_4$ & 215.220 & 44.1 & & 1.0 \\
SO $5_6-4_5$ & 219.949 & 35.0 & & 1.0 \\
SO $2_1-1_2$ & 236.452 & 15.8 & HC$_3$N $26-25$ & 0.0 \\
SO $3_2-2_3$ & 246.404 & 21.1 & several & 0.0 \\
SO $6_5-5_4$ & 251.825 & 50.7 & CH$_3$OH $6_33-6_24$ & (1.0,0.98,0.82,0.97,1.0) \\
 & . & & CH$_3$OH $4_32-4_23$ & (0.84,0.55,0.87,0.81,0.65) \\
SO $9_8-8_8$ & 254.573 & 99.7 & & 1.0 \\
SO $6_6-5_5$ & 258.256 & 56.5 & HC$^{15}$N $3-2$ & 1.0 \\
SO $6_7-5_6$ & 261.844 & 47.6 & CH$_3$OH $2_11-1_01$ & (0.88,0.76,0.74,0.82,0.89) \\
 & . & & C$_2$H $3_4-2_3$ & (1.0,1.0.0.91,1.0,1.0) \\
SO $4_3-3_4$ & 267.198 & 28.7 & HCN,v2=1 $3-2$ & 0.0 \\
SO $1_1-0_1$ & 286.340 & 15.2 & H$_2$C$^{18}$O $4_13-3_12$ & 1.0 \\
SO $5_4-4_5$ & 294.768 & 38.6 & NH$_2$CHO $14_2,13-13_2,12$ & 0.0 \\
SO $10_9-9_9$ & 295.356 & 120.2 & H$_2^{34}$S $3_30-3_21$ & 0.0 \\
SO $7_6-6_5$ & 296.550 & 64.9 & HCNH$^+$ $4-3$ & (1.0,1.0,0.86,1.0,1.0) \\
SO $7_7-6_6$ & 301.286 & 71.0 & & 1.0 \\
SO $7_8-6_7$ & 304.078 & 61.1 & SiO $7-6$ & 1.0 \\
 & . & & OCS $25-24$ & (1.0,1.0,0.87,1.0,1.0) \\
 & . & & CH$_3$OH $2_11-2_02$ & (1.0,1.0,0.96,1.0,1.0) \\
SO $2_2-1_2$ & 309.502 & 19.3 & & 1.0 \\
SO $2_1-1_0$ & 329.385 & 15.8 & C$^{18}$O $3-2$ & 0.0 \\
SO $11_{10}-10_{10}$ & 336.554 & 142.8 & HC$_3$N $37-36$ & 0.0 \\
SO $3_3-2_3$ & 339.341 & 25.5 & & 1.0 \\
SO $8_7-7_6$ & 340.714 & 81.2 & HC$^{18}$O$^+$ $4-3$ & (0.92,0.87,0.93,0.85,1.0) \\
SO $8_8-7_7$ & 344.311 & 87.5 & HC$^{15}$N $4-3$ & (1.0,0.99,0.94,0.96,1.0) \\
SO $3_2-1_2$ & 345.705 & 21.1 & CO $3-2$ & 0.0 \\
SO $8_9-7_8$ & 346.528 & 78.8 & HC$_3$N,v7=1 $38-37$ & (1.0,1.0,1.0,0.90,1.0) \\
 & . & & SO$_2$ $19_1,19-18_0,18$ & (1.0,1.0,1.0,0.98,1.0) \\
SO $7_6-6_7$ & 361.351 & 64.9 & & 1.0 
\enddata
\tablenotetext{a}{SO energy levels designated using the N$_J$ notation.}
\tablenotetext{b}{Overlap correction key: GMCs (3,4,5,6,7), except if all are 0.0 or 1.0.}
\end{deluxetable*}

\subsection{Additional ALMA Archival Data}
\label{sec:pwater3020}
Only two transitions of \pwater{} are contained in the ALCHEMI data. This may limit our ability to constrain our fits and therefore we augmented the ALCHEMI data analyzed with ALMA archival measurements of the \pwater{} $3_0-2_0$ transition. This transition has a rest frequency of 396.272\,GHz and the data was taken from ALMA project 2016.1.01285.S (PI: Jesus Mart\'in-Pintado). These data include imaging of a single primary beam (PB) toward phase center position $\alpha=00^h47^m33.134^s$,  $\delta= -25^{\circ}17'19.68''$ (ICRS) with the ACA ($\theta_{PB} = 38$\arcsec) and 12m Array ($\theta_{PB} = 22$\arcsec).  The observed phase center position for these measurements is within 2\arcsec of the observed phase center for the ALCHEMI measurements (Section \ref{sec:alchemidata}). In order to directly compare these \pwater{} $3_0-2_0$ measurements with our other \pwater{} and SO measurements extracted from the ALCHEMI archive, we have imaged the \pwater{} $3_0-2_0$ observations using the ALCHEMI imaging pipeline \citep{Martin2021}.  Starting with the ALMA-calibrated measurement set from the ALMA archive, we produced continuum-subtracted \pwater{} $3_0-2_0$ image cubes using a robust parameter of 0.5. To match the spatial and spectral resolution of the ALCHEMI measurements, the \pwater{} $3_0-2_0$ images were imaged to a spatial and spectral resolution of 1.6\arcsec\ and 10\,km s$^{-1}$, respectively.  The final spectral channel RMS of these images is 2.6\,mJy beam$^{-1}$. The ALMA recommendation for $1\sigma$ absolute flux calibration uncertainty for measurements from ALMA Band 8 is 20\% (ALMA Cycle 4 Proposer's Guide, Section A.9.2), a value which we adopt for the subsequent analysis of these \pwater{} $3_0-2_0$ measurements.  We have also extracted the continuum emission associated with these measurements using the continuum subtraction and imaging process described in \citet{Martin2021}.
\subsection{Spectral Line Overlap}
Local thermodynamic equilibrium (LTE) modelling of the ALCHEMI data \citep{Martin2021} showed that the moderate line widths ($\Delta$v $\simeq 75$\,km/s FWHM) of the transitions detected towards \gal{} cause many transitions to overlap. As a result, measuring the emission above the noise in the vicinity of each of our transitions will often result in total integrated intensities that include flux from other transitions.\par
In order to estimate which of our SO and \pwater{} transitions are expected to be significantly blended with an interloper, we used the modeling results described by \citet{Martin2021} as a guide, with inspection of the spectra from each GMC. Where a target SO or \pwater{} transition was found to be potentially blended with the emission from another line, we estimate the amount of overlap between the integrated intensities of these two transitions.  We then take the integrated intensity to be only the fraction of the emission which comes from the target transition.\par
\begin{figure}[!h]
\includegraphics[width=0.5\textwidth]{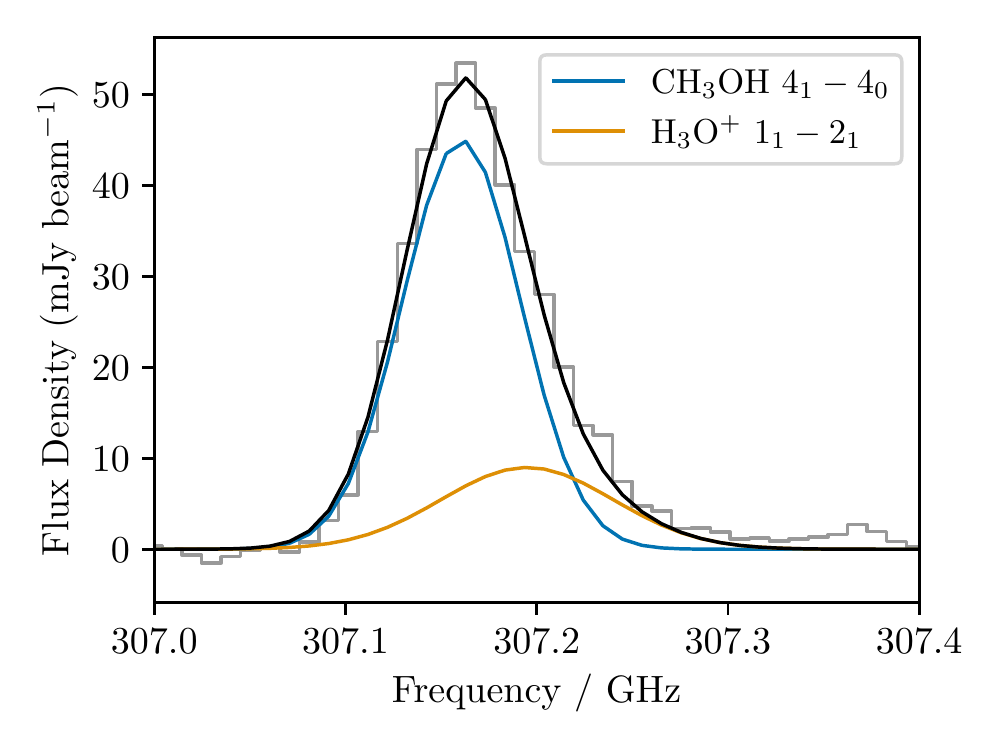}
\caption{Observed spectrum from GMC 7 covering the \pwater{} 1$_1$--2$_1$ transition in grey with a multiple gaussian fit plotted in black. The colour traces show the individual Gaussians used in the fit labelled by the transition which corresponds to the central frequency.  The derived interloper correction factors, using Equation~\ref{eq:interlopercorrection}, are listed in Table~\ref{table:transitions}.}
\label{fig:interloperexample}
\end{figure}

The overlap estimate is derived by performing multiple Gaussian fits to spectra drawn through beam-sized areas centered on the position of each GMC. The interloper overlap correction using these sample spectra is defined as
\begin{equation}
    \mathrm{Interloper~Correction} \equiv 1-\frac{\sum_{overlap} y_{interloper}}{\sum_{overlap} (y_{target} + y_{interloper})}
    \label{eq:interlopercorrection}
\end{equation}
where $y_{interloper}$ and $y_{target}$ are the intensities of the relative frequency-constrained Gaussian fits to the target and interloper transitions, respectively, within a spectral channel. The sum $\Sigma_{overlap}$ is taken over all spectral channels with signal larger than the RMS noise in the image cube under consideration.  An example of the Gaussian fit analysis used to derive the interloper correction factors defined by Equation~\ref{eq:interlopercorrection} and listed in Table~\ref{table:transitions} is shown in Fig.~\ref{fig:interloperexample}.
As is apparent from the interloper correction factors listed in Table~\ref{table:transitions} and the spectrum shown in Figure~\ref{fig:interloperexample}, the overlap between the \pwater{} $1_1-2_1$ and CH$_3$OH $4_1-4_0$ transition is quite large, as the two transitions are separated by just 26\,MHz ($\sim 2.5$ spectral channels).  As the \pwater{} $1_1-2_1$ transition is one of only three \pwater{} transitions available in this study, it is important to properly assess the quality of our \pwater{} spectral line extraction for this transition.  
The contribution from CH$_3$OH to the observed spectral profile was derived from LTE modelling using MADCUBA \citep{Martin2019} to both the CH$_3$OH molecular emission across the entire ALCHEMI frequency range as well as using only nearby surrounding (in frequency and energies) CH$_3$OH transitions.  This LTE modeling procedure resulted in a derived \pwater{} to CH$_3$OH spectral line peak intensity ratio of 0.2.  Using the procedure described above, with this derived Gaussian-peak relative intensity, we derive the overlap corrections listed in Table~\ref{table:transitions}.\par
\subsection{Spectral Line Moment Extraction}
\label{momentanal}
Once all spectral interlopers were identified and their influence on the spectrally-integrated emission from our target SO and \pwater{} transitions assessed, we proceeded to calculate the zeroth, first, and second moments of all detected SO and \pwater{} transitions using the \texttt{CubeLineMoment}\footnote{\url{https://github.com/keflavich/cube-line-extractor}} script introduced for this same purpose by \cite{Mangum2019}.  \texttt{CubeLineMoment} uses a series of spectral and spatial masks to extract integrated intensities for a defined list of target spectral frequencies.  As noted by \cite{Mangum2019}, the \texttt{CubeLineMoment} masking process uses a bright spectral line whose velocity structure is representative of the emission over the galaxy as a ''tracer'' of the gas under study. In almost all of the SO and \pwater{} transitions studied in this analysis, we use the target transition itself as the tracer transition as our target transitions are generally sufficiently intense. Final zeroth, first, and second moment images were generated using a signal limit of three times that of the spectral channel baseline rms for the respective transition under study.  Figure~\ref{fig:H3Op320221IntSO3423Int} shows representative samples of the \pwater{} and SO integrated intensity (zeroth moment) images resultant from this analysis.  In Appendix~\ref{sec:AppendixIntFigs} we show the remaining \pwater{} integrated intensity images and samples of SO emission from the other ALMA receiver bands from our measurements (Bands 3, 5, 6, and 7).
\par
As a result of this extraction, we obtain the integrated line intensity of every transition in Table~\ref{table:transitions} for each GMC in the CMZ. This is done by taking the average intensity over a beam sized region centred on the GMC centres defined by \citet{Leroy2015}. We then adopt uncertainties on these intensities which combine the recommended 15\% (for ALCHEMI measurements) or 20\% (for \pwater{} $3_0-2_0$ measurements) absolute calibration uncertainty and the integrated noise added in quadrature. Finally, we adjust each integrated intensity by the relevant interloper correction factor given in Table~\ref{table:transitions}. We do not adjust the uncertainty of a transition, choosing instead to retain the total uncertainty on the blended emission. This effectively overestimates the uncertainty on the intensity of the transition to account for our uncertainty on the interloper correction factor. All integrated intensities, uncertainties and other relevant transition properties are uploaded in a supplementary table.
\section{Modelling Approach}
\label{sec:approach}
\subsection{The Forward Model}
\label{sec:model}
To infer the cosmic ray ionization rate, we require a model which can take a small number of free parameters, including the cosmic ray ionization rate, and ultimately return line intensities for all the detected lines of \pwater{} and SO. To do this, we combine the gas-grain chemical model UCLCHEM \citep{Holdship2017UCLCHEM} with RADEX \citep{VanderTak2007} which is used via the Python package SpectralRadex\footnote{\url{spectralradex.readthedocs.io}}. We use recently published collisional rates between \pwater{} and p-H$_2$ \citep{Demes2021} as well as SO collisional data \citep{Lique2006}, both taken from the LAMDA\footnote{\url{https://home.strw.leidenuniv.nl/~moldata/}} database \citep{Schoier2005LAMDA} .\par
In the chemical models, we assume the GMCs can be modelled as uniform clouds with fixed physical conditions and a visual extinction that is sufficiently high to make photo-processes negligible. This simple model is justified in the next section. The simplified picture allows us to use a single point model as the only depth dependent effect in UCLCHEM is the UV attenuation which becomes unimportant if the majority of the gas considered is at high visual extinction. This is important as even a 1D modelling approach would be computationally unfeasible when combined with extensive sampling of the necessary parameter space. One caveat of this model is that it is known that the CRIR is attenuated by column density \citep{Padovani2018}. However, that attenuation is weak for the column densities considered in this work and so inferring an average value is reasonable. \par
The above chemical modelling requires three main inputs: the gas volume density ($n_{H2}$), the gas kinetic temperature ($T_{kin}$) and the cosmic ray ionizaton rate ($\zeta$) which will be given in units of $\zeta_0$=\SI{1.36e-17}{\per\second} throughout this work as that is the normalization factor used in the chemical network \citep{McElroy2013}. In addition, we assume the initial elemental abundances are the depleted values from \citet{Jenkins2009} or that they are scaled from those values by a constant metallicity factor ($Z$). The exception to this is sulfur for which we treat the elemental abundance at $Z=1$ as a free parameter (see Sect.~\ref{sec:inference}). The model returns the equilibrium abundances of \pwater{} and SO. By adding the additional parameters of the total H$_2$ column density (N$_{H2}$) of the GMC and the linewidth ($\Delta V$), we can generate line intensities for all transitions of these two species by assuming a spherical GMC and utilizing RADEX. RADEX requires the column density of each species which is obtained by multiplying the fractional abundance from the chemical model by the H$_2$ column density. The collisional excitation rate file for \pwater only contains rates for collisions with p-H$_2$.  For this reason we set the p-H$_2$ density in RADEX to $n_{H2}$ for \pwater calculations on the grounds that assuming the o-H$_2$ collisional rates are equal to the p-H$_2$ rates must be a better approximation than assuming no collisions with o-H$_2$.\par
\subsection{Model Justification}
\label{sec:prelim}
The GMCs in NGC253 are complex and many physical processes are at play. Thus, two things must be demonstrated to justify the use of this modelling procedure to infer the cosmic ray ionization rate. First, we must show the forward model and, in particular, the \pwater{} to SO ratio is sensitive to the cosmic ray ionization rate as suggested by \citet{Bayet2011}. Second, we must show that other physical or chemical processes can be neglected. At a minimum, the outer regions of any dense clouds that compose the single objects we observe at a resolution of 28 pc will be UV irradiated.  Further, on larger angular scales ($\gtrsim 15$\arcsec) low-velocity shocks have been shown to be a dominant heating mechanism in the \gal{} CMZ \citep{Martin2006} and so shock chemistry may be at work. Thus our simple, single point model with negligible influence from an external UV field requires justification.\par
\begin{figure*}
    \centering
    \includegraphics[width=\textwidth]{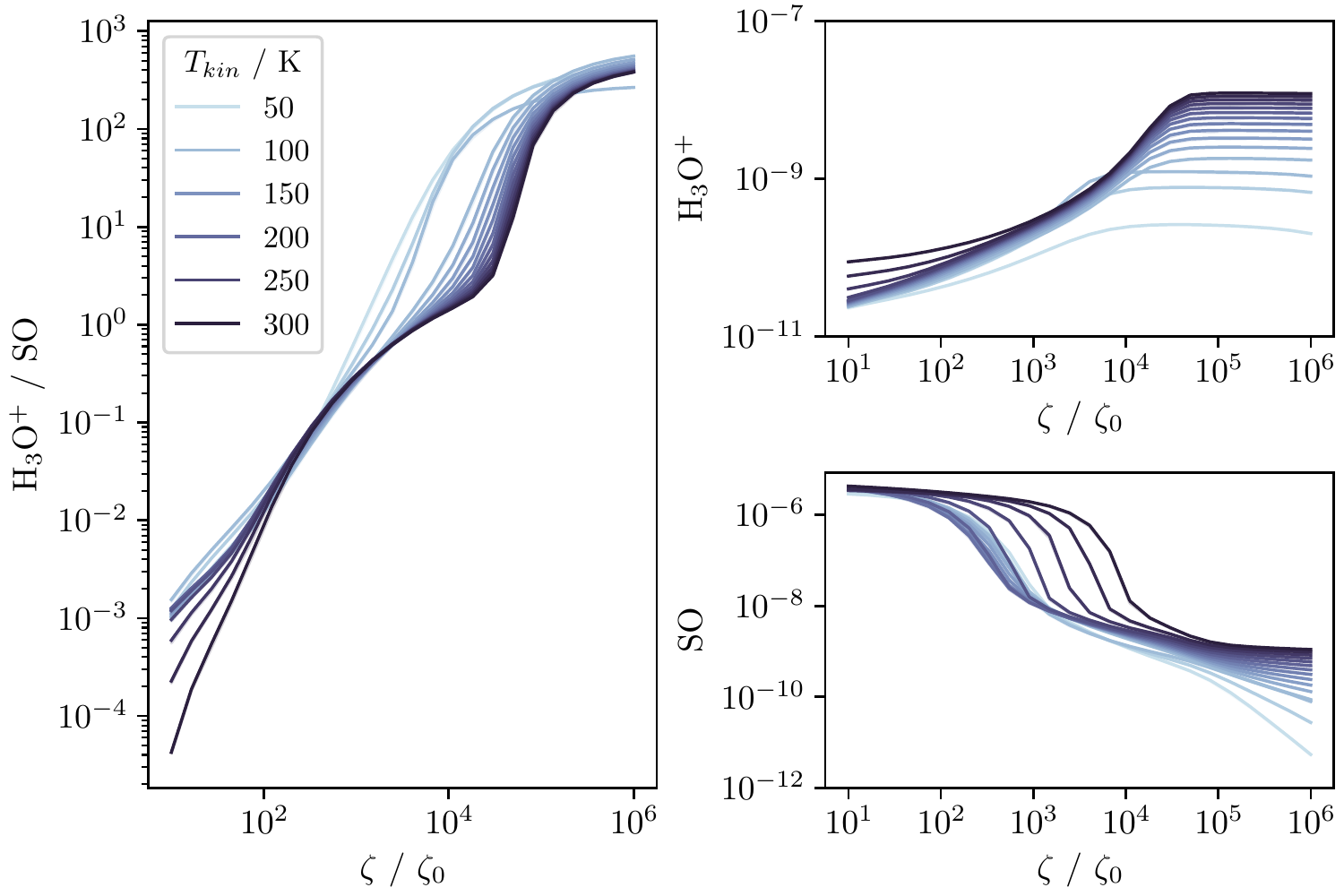}
    \caption{Equilibrium \pwater\ and SO abundances and their ratio as a function of cosmic ray ionization rate given in units of $\zeta_0$=\SI{1.3e-17}{\per\second}. The abundances are averaged over models with different densities in the range \num{e4} -- \SI{e7}{\per\centi\metre\cubed} and the colour of the lines indicates the gas temperature.}
    \label{fig:ratiotest}
\end{figure*}

To address the model dependence on $\zeta$, a grid of UCLCHEM models were generated in which the density, temperature and cosmic ray ionization rate of a cloud were varied assuming a large visual extinction. Figure~\ref{fig:ratiotest} shows the steady state abundances of \pwater\ and SO across the models of this grid as well as their ratio. Both species show a clear and strong dependence on the cosmic ray ionization rate with the \pwater\ abundance increasing and the SO abundance decreasing in response to an increasing $\zeta$.\par
In the case of \pwater{} this is driven by the fact that \pwater{} is formed through the following chain of reactions,
\begin{eqnarray}
     {\rm H_2 + OH^+ \longrightarrow H_2O^+ +H} \\
    {\rm H_2 + H_2O^+ \longrightarrow  H_3O^+ + H}\nonumber
\end{eqnarray}
where each reaction is the primary destruction route of the ionic reactant. Therefore, the \pwater{} abundance primarily depends on the OH$^+$ abundance which is primarily produced from both reactions between O$^+$ and H$_2$ and between H$^+$ and OH. In both cases, the ions are formed directly by cosmic rays and thus increasing $\zeta$ generally increases the overall \pwater{} production rate. This ceases to be the case once $\zeta$ is such that the ionization fraction of the gas is very high. At that point, it has been shown that increasing dissociative reactions with electrons actually drive a decrease in the \pwater abundance \citep{Gerin2010InterstellarG10.6-0.4}. This is most noticable in the T$_{kin}$ = 50 K case in Figure~\ref{fig:ratiotest} where the H$_3$O$^+$ abundance actually starts to decrease at the highest $\zeta$.\par
SO is a much simpler case. It forms efficiently through neutral-neutral and ion-neutral reactions in the gas-phase, obtaining relatively high abundances ($\sim$ \num{e-6}) in most gas conditions. However, its primary destruction routes are reactions with ions and cosmic rays, the majority being destroyed in reactions with C$^+$ and H$^+$. Since the abundance of these reactants is directly tied to the CRIR, SO is destroyed more efficiently as $\zeta$ increases. Fortunately, this appears to be even more efficient at low temperatures, when the \pwater abundance is least sensitive to $\zeta$.\par
The opposite responses of these species to the CRIR results in their ratio being highly sensitive to the value of $\zeta$. In fact, it varies by seven orders of magnitude over the explored range of $\zeta$. Moreover, the variation in the ratio due to $\zeta$ is much larger than the variation due to temperature, so uncertainty in the gas temperature will not prevent us from inferring the cosmic ray ionization rate. A final point to note about this preliminary modelling is that the steady state abundances are reached quickly - typically within \num{e5} years. This justifies the use of equilibrium abundances for the GMCs.\par
\begin{figure*}
    \centering
    \includegraphics[width=\textwidth]{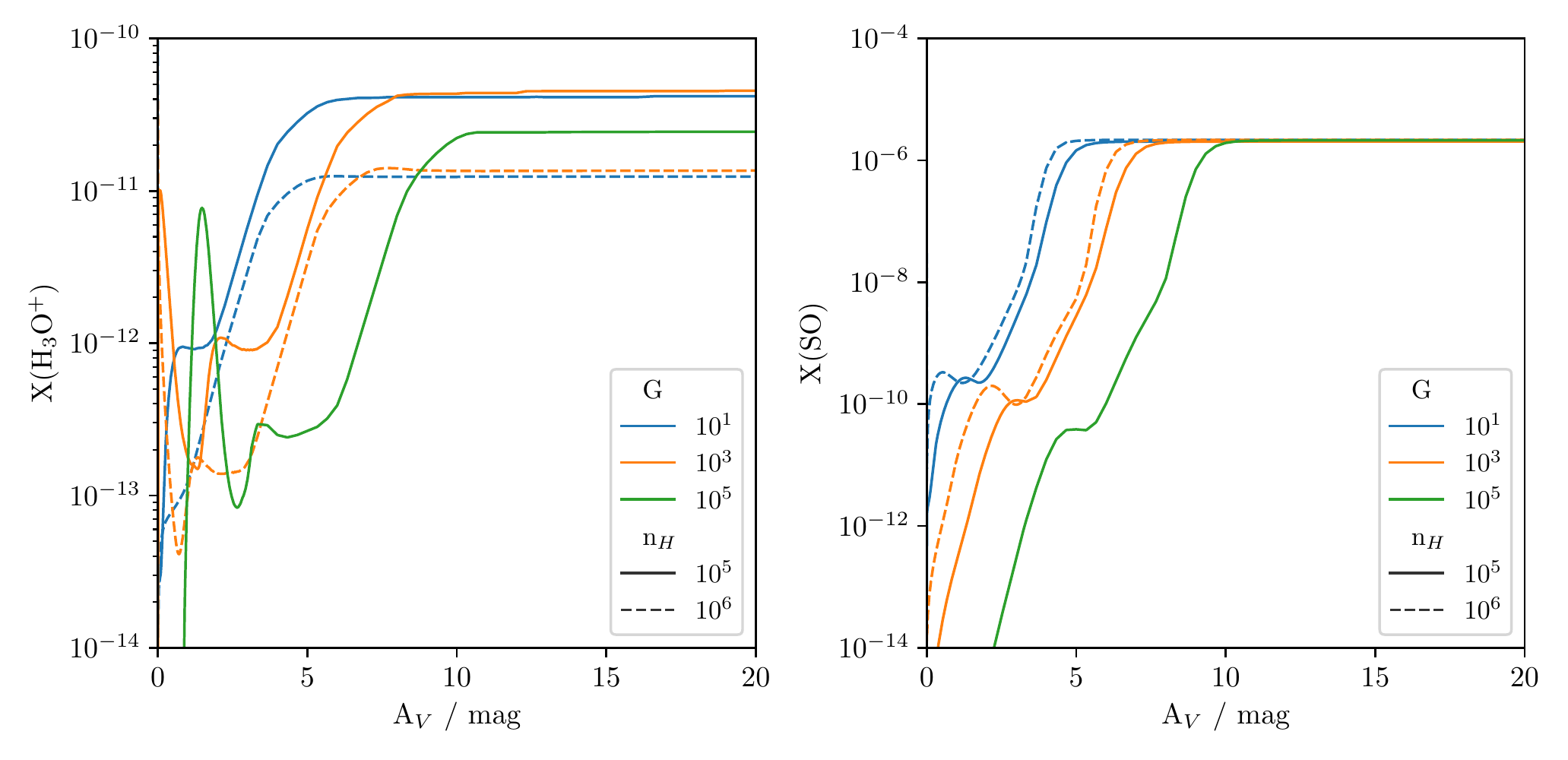}
    \caption{Abundance as a function of visual extinction for \pwater and SO from UCL\_PDR models. The majority of these species come from within the high A$_V$ parts of the cloud.  This is particularly true for SO, which is as much as eight orders of magnitude higher in abundance once the $A_V$ is sufficiently high to reduce photo-proccesses to zero than it is at the cloud edge.}
    \label{fig:pdr_model}
\end{figure*}

To determine whether UV processes can be neglected, we use UCL\_PDR\footnote{\url{https://github.io/UCL_PDR}} \citep{Bell2006,Priestley2017} to determine whether the assumption that \pwater\ and SO primarily arise from gas where the visual extinction is high is appropriate. This is a 1D model which solves the equilibrium temperature and abundances for a semi-infinite slab of gas. Fig.~\ref{fig:pdr_model} demonstrates that for a broad range of conditions which are reasonable for the GMCs under study, the vast majority of the \pwater\ and SO columns in these PDR models arise from deeper within the model cloud. Thus any \pwater\ and SO emission from these objects should primarily trace the higher column density regions where the visual extinction is greater than 5 magnitudes.\par
Finally, the possibility that shocks throughout the gas affect the ratio of \pwater and SO can also be addressed through preliminary modelling. To evaluate this scenario, we generated a range of shock models using the C-shock parameterization of \citet{jimenez2008} which includes sputtering of the ice mantles. We consider shock velocities between 5 and \SI{40}{\kilo\metre\per\second} through gas with a preshock density of \num{e5} and \SI{e6}{\per\centi\metre\cubed}. We find that whilst the shock passage does tend to enhance the abundances of both species, the ratio of the two species is relatively constant at  \pwater/SO $\sim$ \num{e-4} in shocked gas. Thus it will be important to test whether the abundance ratios we obtain are close to this value but otherwise, we can assume the ratio is not dominated by shocks.\par
In summary, our preliminary modelling shows that the ratio of \pwater\ and SO is strongly dependent on the ionization rate of the gas and comparatively weakly dependent on the temperature. Furthermore, SO almost entirely arises from the inner regions of clouds with high visual extinction and the \pwater\ abundance is also higher in these regions than close to the cloud edge so neglecting the UV processes is justified. It is likely there are shocks present in the region under study, but they tend to produce a constant \pwater\ to SO ratio.  If the chemistry of these species is shock dominated, this can be checked \textit{a posteriori} by evaluating the \pwater\ to SO ratio found by the inference.
\subsection{Parameter Inference}
\label{sec:inference}
With the model described in Section~\ref{sec:model}, we can use Bayesian inference to find the probability distribution of the values of the model parameters given the data that we observed (Section~\ref{sec:data}). From Bayes' theorem, we know this probability distribution is given by,
\begin{equation}
    p(\theta | d) = \frac{\mathcal{L}( d | \theta)p(\theta)}{p(d)}
    \label{eq:bayes}
\end{equation}
where $\theta$ represents the parameters and $d$ the data. The likelihood ($\mathcal{L}( d | \theta)$) is easy to formulate as our data have normally distributed uncertainties, which allows us to use the standard Gaussian likelihood,
\begin{equation}
    \mathcal{L}( d | \theta)= \exp\left(-\frac{1}{2} \sum_i{\frac{(d_i-M_i)^2}{\sigma_i^2}} \right)
\end{equation}
where $d_i$ is the measured intensity of one transition, $\sigma_i$ is the uncertainty of that measurement, and $M_i$ is the model spectral line intensity for a given model parameter $\theta$.\par
The evidence ($p(d)$) is a constant for all $\theta$ for a given model and so we neglect it in this work. To obtain the final probability distribution $p(\theta|d)$, the probability distribution obtained from the numerator of Equation~\ref{eq:bayes} is simply normalized.\par
Finally, a prior distribution $p(\theta)$ must be chosen. The large amount of previous work on \gal{} provides a fantastic opportunity to set informed priors on the density \citep{Leroy2018}, column density \citep{Mangum2019}, and CRIR \citep{Holdship2021,Harada2021}. Despite this, we chose in the first instance to use uninformative priors which are uniform within limits based on those works. Finding that we adequately constrained our parameters with these uninformative priors and the data at hand, we did not progress to using informed priors. The ranges of the uniform priors are given in Table~\ref{table:priors}.\par
A Bayesian inference approach gives us a simple way to deal with several unknown parameters. In Table~\ref{table:priors} we list the parameters of interest ($n_H, T_{kin}, N_{H2}, \zeta$) as well as several nuisance parameters ($\Delta V, S, Z, o/p$) which are unknown parameters that are not of interest to our work but may affect our model fits. For example, it is well known that in Galactic environments as little as 1\% of sulfur is accounted for in dense gas \citep{charnley1997}. Thus it is unclear how much sulfur will be available for gas-phase reactions in \gal{} and so we leave the elemental abundance of sulfur at a metallicity of 1 as a free parameter (S). Including nuisance parameters in this way will account for the increased uncertainty in our inferred values for the parameters of interest due to our lack of knowledge.\par
In addition to the sulfur depletion factor, we consider that the exact metallicity (Z) of \gal{} is unknown but is similar to the solar metallicity \citep{Marble2010} and so allow it to vary within a small range. We also require an ortho:para ratio of \pwater\ which UCLCHEM does not provide. Given that this ratio has limiting values of 2 at temperatures below \SI{50}{\kelvin} and 1 at temperatures above \SI{100}{\kelvin}, we allow the ratio to vary freely in this range.\par
\begin{deluxetable}{lccc}
    \tablewidth{0pt}
    \tablecaption{Prior distributions used for each parameter\tablenotemark{a}}
    \label{table:priors}
    \tablehead{
        \colhead{Symbol} & \colhead{Name} & \colhead{Prior Type} & \colhead{Range}
    }
    \startdata
         $n_H$& Gas density & log-uniform & \num{e4}--\SI{e7}{\per\centi\metre\cubed} \\
         $T_{kin}$ & Gas temperature & uniform & 50--300 K\\
         $N_{H2}$ & H$_2$ column density& log-uniform & \num{3e22}--\SI{e25}{\per\centi\metre\squared}\\
         $\zeta$ & CRIR & log-uniform & \num{e3}--\num{e7} $\zeta_0$\\
         $\Delta V$ &Line width& uniform & 50--\SI{150}{\kilo\metre\per\second} \\
         S & S abundance& log-uniform & 0.01 -1 S$_{\odot}$\\
         Z & Metallicity & uniform & 0.5--3 Z$_{\odot}$\\
         $o/p$ & \pwater ortho:para ratio & uniform & 1-2\\
    \enddata
    \tablenotetext{a}{S $\equiv$ elemental sulfur abundance at Z=1 with $S_{\odot}$ its solar value.}
\end{deluxetable}
%
%
Finally, in order to obtain the probability distribution $P(\theta|d)$, the right hand side of Equation~\ref{eq:bayes} was sampled using \texttt{Ultranest} \citep{Buchner2021} which is a nested sampling package. Given the relatively computationally intensive nature of the forward model, the parameters utilized in UCLCHEM were limited in precision by rounding. The temperature was rounded to the nearest \SI{1}{\kelvin} and other UCLCHEM input parameters were rounded to the nearest 0.1 dex. This allowed model results to be saved and then if very similar parameters were requested by the sampler, the abundances could be read from file rather than repeatedly running UCLCHEM for very small changes in parameter values. As a result the posterior distribution of $\theta$ is limited in precision to the same degree.\par
\section{Results}
\label{sec:results}

\subsection{Model Spectral Line Intensities}
\begin{figure}
    \includegraphics[width=0.5\textwidth]{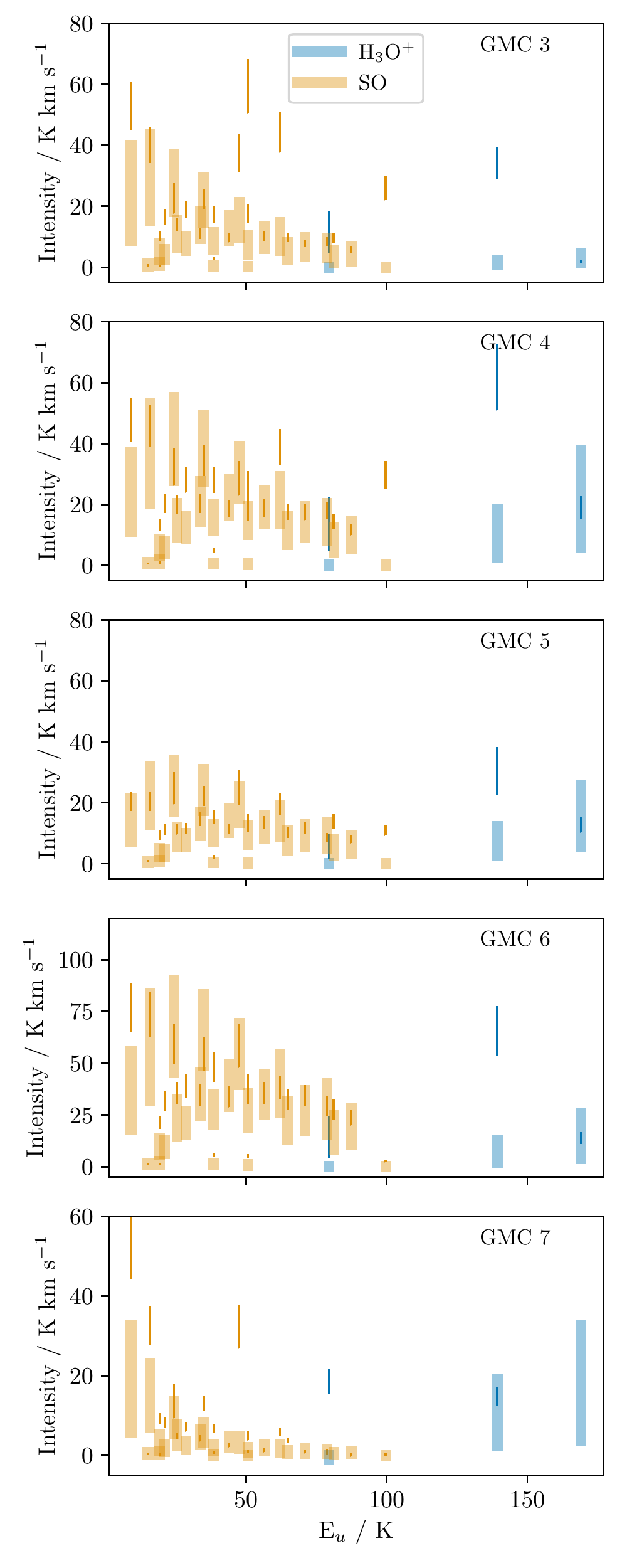}
    \caption{Measured line intensities for the GMCs plotted as error bars which cover the 1$\sigma$ uncertainty range. Shaded bars show the 16th to 83rd percentile of fluxes predicted by our model when sampling from the posterior distribution.}
    \label{fig:fluxplots}
\end{figure}

\texttt{Ultranest} provides a set of samples from the posterior distribution weighted so that points appear in proportion to how likely they are. We use this to evaluate the goodness of fit of our model by generating the line fluxes using all points in this sample and then comparing them to the measured fluxes. The observed intensities with 1$\sigma$ errorbars are plotted in Figure~\ref{fig:fluxplots} alongside the 16th to 83rd percentile range of the posterior sample fluxes. This range was chosen because for a Gaussian distribution it is equivalent to the 1$\sigma$ range.\par
%
Ignoring clear outliers, the majority of transitions show an overlap between the measurement uncertainties and the range of fluxes predicted by the model. Some outlying transitions are to be expected considering that we are using a highly simplified model of a homogeneous sphere to fit complex regions with varying densities and temperatures. Points that do not overlap tend to be low E$_U$ SO transitions which are underpredicted by the model. This is perhaps a similar effect to that seen in C$_2$H emission in the same GMCs \citep{Holdship2021} where low E$_U$ emission was clearly excited by an additional gas component. However, since these deficiencies are small, we consider that the majority of the gas emitting these species is well characterized by our model.\par
The \pwater\ transition at E$_u$ = \SI{79,5}{\kelvin} ($1_1-2_1$ near 307\,GHz) is also often underfit. However, this is the transition which suffers from a large degree of overlap with a much stronger CH$_3$OH line as shown in Figure~\ref{fig:interloperexample}. It is possible that the overlap correction factors are too large and we are assigning CH$_3$OH flux to the \pwater\ line. However, the \pwater\ transition at E$_u$ = \SI{129.8}{\kelvin} ($3_2-2_2$ near 365\,GHz) is also much stronger than our models predict. No potential interloper is strong enough to explain the additional flux, but it is possible non-thermal excitation effects are contributing to the excitation of this transition. Under certain conditions, this transition can be infra-red pumped \citep{Phillips1992,Martin2021} which would explain the additional flux. \par
We checked that these outliers do not bias our results by fitting the GMCs  excluding the 307 and 365 GHz transitions.  The H3Op abundance in these models is then constrained by just the 396 GHz transition, which is not expected to be radiatively pumped \citep{Phillips1992}. We find that we obtain similar results, indicating these outliers are not strongly biasing our fit and it is primarily the fit to SO that is driving our results.\par
It is possible that the chemistry creates a complicating factor which prevents us finding adequate column densities to fit the data or to simultaneously fit both species. We rule this out by fitting the data with RADEX only, treating the SO and \pwater{} column densities as free parameters alongside the gas temperature and density. We find the same best fit fluxes are obtained, indicating that no simple RADEX model can fit these data and it is not the case that we simply cannot obtain the required abundances in our chemical model.\par
Overall, whilst a better radiative transfer model, or perhaps simply a multiple gas component fit to the data, would likely give better results, we consider that our simple model fits are sufficient. Moreover, as discussed in Sect.~\ref{sec:gmc-props}, the inference of the physical parameters allows us to validate our results using previous measurements from each GMC modelled.\par
\subsection{Posterior Distributions}
The specific values of each posterior distribution differ from GMC to GMC but the general trends are similar.  We discuss here the corner plot for GMC 4 shown in Fig.~\ref{fig:corner4} and include the corner plots for the other GMCs modelled in Appendix~\ref{sec:corners} for completeness.\par
\begin{figure*}
    \centering
    \includegraphics[width=\textwidth]{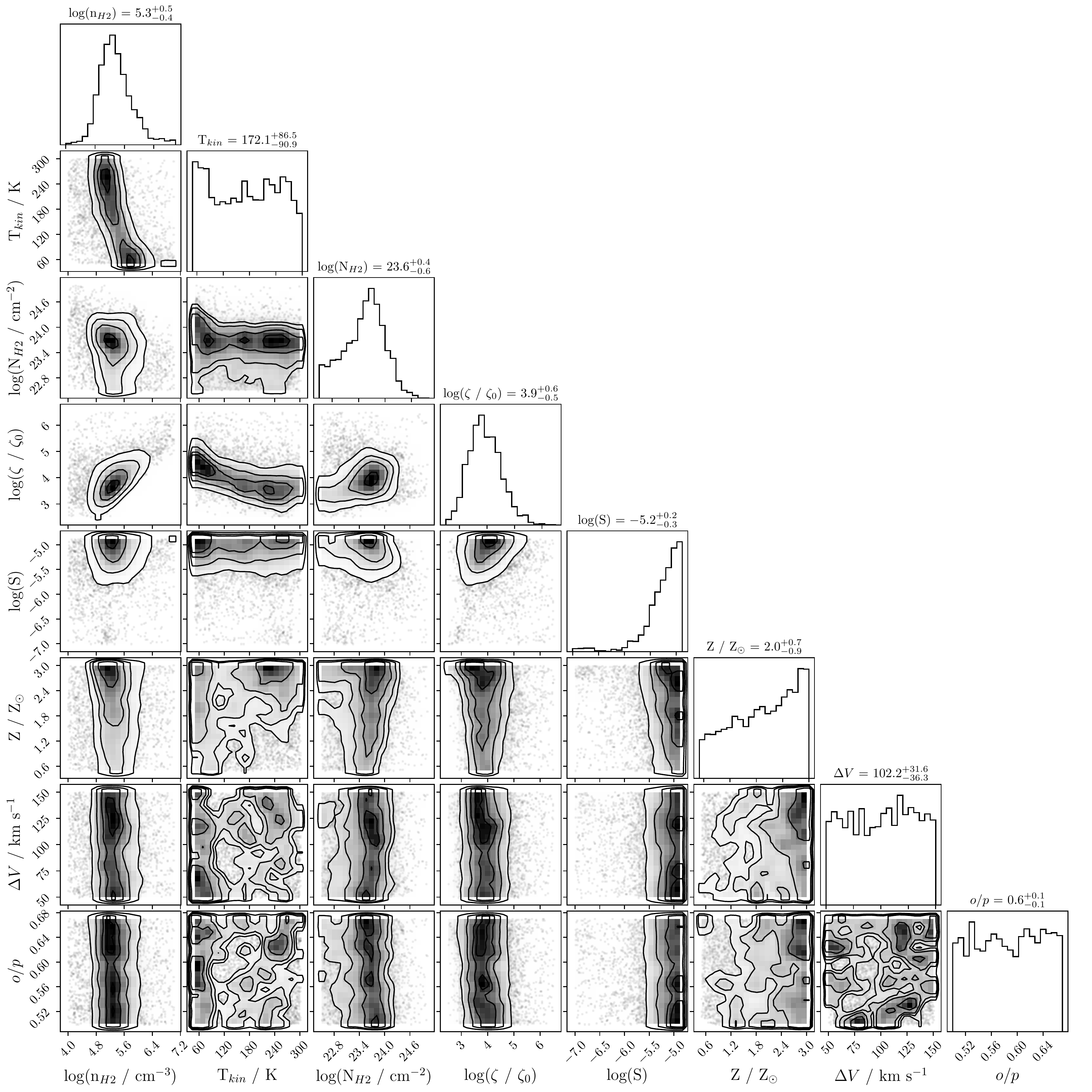}
    \caption{Corner plot showing the marginalized posterior probability distribution of every parameter and the joint distributions of all parameter pairs for GMC 4. Note the ortho:para ratio of \pwater is given as the fraction of the \pwater in ortho form.}
    \label{fig:corner4}
\end{figure*}
From the 1D marginalized posterior distributions, it is clear that we constrain three of our four parameters of interest. The gas density, the H$_2$ column density, and the cosmic ray ionization rate are well constrained whilst the temperature is not. Each of the three constrained parameters has a strong peak indicating a most likely value though the H$_2$ column density peak becomes flat below a certain limit. This is due to the fact that below a certain value all column densities produce intensities well below the rms noise in our measurements, and thus give identically poor fits.\par
The 2D joint posteriors show some interesting degeneracies. In particular, the gas density and cosmic ray ionization rate shows a log-log relationship which arises from a degeneracy in our chemical model.  In UCLCHEM, the abundance can remain unchanged for increasing CRIR if the density also increases. However, since the density is also a parameter of RADEX, it is constrained by radiative transfer considerations. Thus, the range of acceptable densities is greatly limited allowing us to determine $\zeta$ despite the chemical degeneracy. Thus, without an accurate measurement of the density, we would not be able to estimate the cosmic ray ionization rate.\par
Finally, there are the four nuisance parameters we included in our inference. The elemental sulfur abundance is constrained but the peak indicates that, whatever the metallicity, the elemental sulfur abundance should have a similar ratio to the other elements as found in the Sun. This means the sulfur cannot be heavily depleted onto the grains in these regions. The other nuisance parameters are unconstrained; the marginalized posteriors are very similar to the priors. This is to the expected result for the line width in the case of optically thin lines. Whilst constraining the metallicity or ortho:para ratio of \pwater{} would have been useful, the nuisance parameters were largely included to account for their effect on the uncertainty of the cosmic ray ionization rate and this is achieved regardless of the fact they are unconstrained. \par
\subsection{The GMC Properties}
\label{sec:gmc-props}
In this section, we present the likely values of the three parameters of interest which we have been able to constrain. Namely, the gas density, the H$_2$ column density and the cosmic ray ionization rate. We present the most likely value of each parameter in each region as well as a most likely interval that contains 67\% of the probability density, similar to a 1$\sigma$ uncertainty. These are given in Table~\ref{table:summary}.\par
\begin{deluxetable*}{rcccccc}
\tablewidth{0pt}
\tablecaption{Most likely values of each well constrained parameter with ranges containing 67\% of the probability density.}
\label{table:summary}
\tablehead{
    \colhead{GMC}&
    \colhead{$n_{H2}$ / \SI{e5}{\per\centi\metre\cubed}}&
    \colhead{$n_{H2}$ Range}&
    \colhead{N$_{H2}$ / \SI{e23}{\per\centi\metre\squared}}&
    \colhead{N$_{H2}$ Range}&
    \colhead{$\zeta$ / \num{e4} $\zeta_0$\tablenotemark{*}}&
    \colhead{$\zeta$ Range}
}
\startdata
 3 & 1.3 & 0.4 - 4.9 & 0.8 & 0.4 - 2.1 & 0.2 & 0.1 - 1.2 \\
 4 & 2.0 & 0.9 - 5.9 & 3.7 & 1.0 - 9.3 & 0.8 & 0.2 - 3.2 \\
 5 & 2.2 & 1.0 - 7.2 & 3.0 & 1.1 - 6.5 & 1.1 & 0.3 - 4.5 \\
 6 & 2.6 & 1.1 - 9.8 & 2.5 & 0.7 - 7.5 & 0.5 & 0.2 - 2.6 \\
 7 & 0.3 & 0.1 - 1.2 & 5.3 & 1.1 - 21.7 & 0.8 & 0.2 - 5.9 \\
\enddata
\tablenotetext{*}{$\zeta_0$ = \SI{1.36e-17}{\per\second}.}
\end{deluxetable*}
The goal of this study is ultimately to infer the cosmic ray ionization rate in the GMCs of \gal{}. It is therefore promising that we have constrained the cosmic ray ionization rate in every case. Each of the five GMCs has a cosmic ray ionization rate of $\zeta \sim$\num{e4} $\zeta_0$. GMC 3 has a most likely value that is a factor of 2-3 lower than the other GMCs and a significant portion of the probability density is at low values. However, it is consistent with the lower end of the range of likely values for the other GMCs and so the difference cannot be said to be significant.\par
\begin{figure*}
    \centering
    \includegraphics[width=\textwidth]{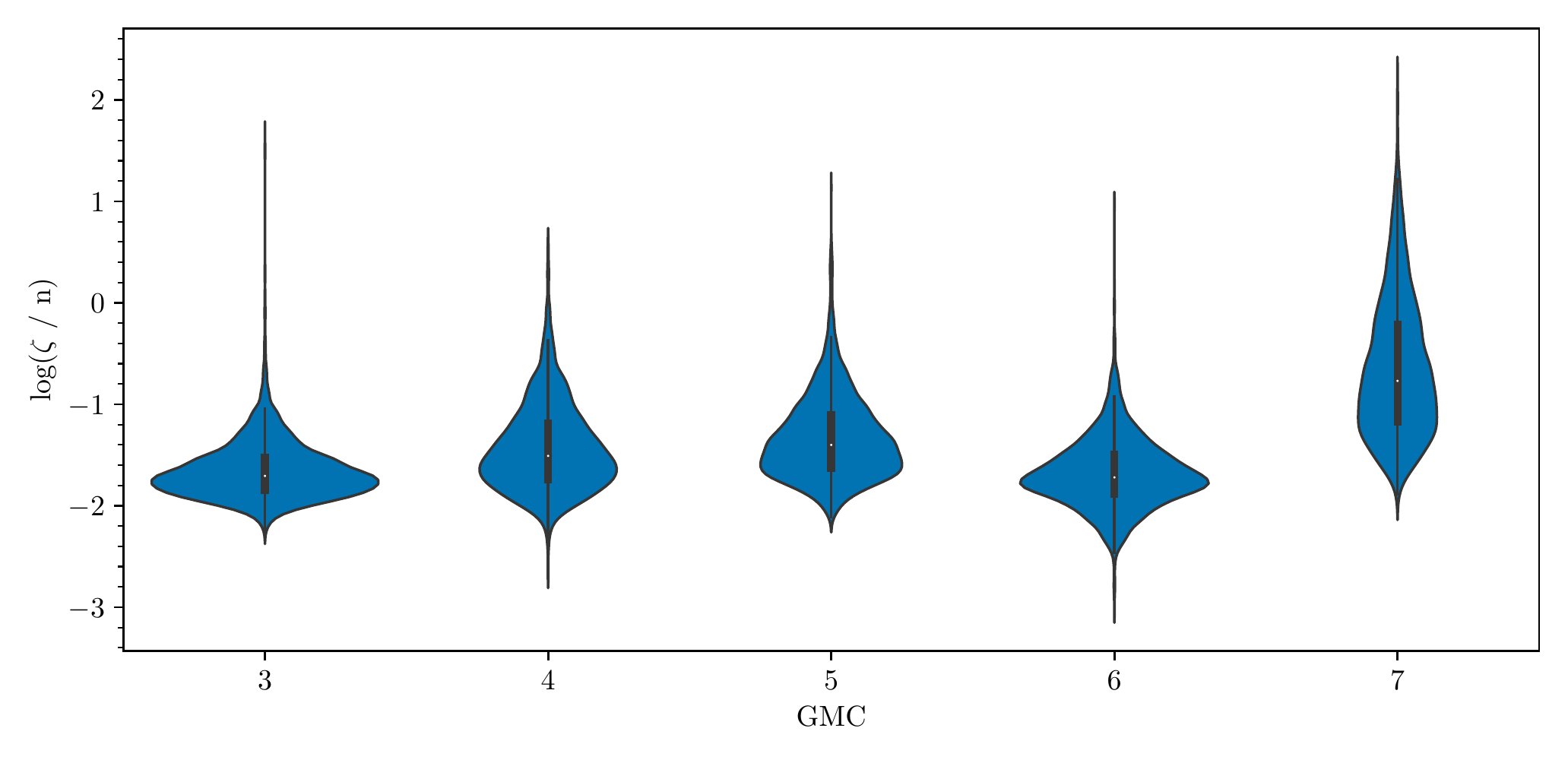}
    \caption{Violin plots showing the posterior probability distribution of $\frac{\zeta}{n}$ in each GMC. The width of the blue area shows the relative probability of a $\frac{\zeta}{n}$ value and the internal box plot marks the quartiles.}
    \label{fig:violins}
\end{figure*}
The column densities we obtain are typically an order of magnitude lower than those obtained from measurements of the dust \citep{Mangum2019}. However, they are consistent with the H$_2$ column densities derived from observations of C$^{18}$O \citep{Harada2021}. This agreement is important as RADEX relies on species column densities. Thus the underlying abundances must be accurate if a good fit is achieved with a reasonable H$_2$ column density. This in turn gives us confidence in our inferred CRIR as we know the SO and \pwater{} abundances are largely set by this parameter.\par
The gas densities are consistent with previous measurements. The range of likely density values for each region overlap very well with those derived from ALCHEMI observations of C$_2$H \citep{Holdship2021}. They also fall well within the range of \num{e5} - \SI{e6}{\per\centi\metre\cubed} found in other ALCHEMI work \citep{Harada2021} and elsewhere \citep{Leroy2018}. Given how strongly correlated $\zeta$ is with this parameter, it is key that these values are accurate. The chemical degeneracy is such that if the density were a factor of 10 lower, our inferred $\zeta$ would be also be a factor of ten smaller. Nevertheless, the good agreement with previously-derived values of the gas number density from the literature should give confidence in our derived CRIR.\par
However, we can also combine the density and CRIR to obtain a posterior distribution on the value of $\frac{\zeta}{n}$. In Figure~\ref{fig:violins}, we show these distributions as violin plots. We actually constrain this value very well indicating that a large part of the uncertainty we report on the CRIR is due to the degeneracy with density. Future use of SO and \pwater as probes of the CRIR could benefit from additional data to constrain the density or stronger priors based on previous observations.\par
\section{Discussion}
\label{sec:discussion}
\subsection{The ALCHEMI view of the CRIR in NGC 253}
We obtain a large cosmic ray ionization rate that is nevertheless consistent with previous measurements. In previous ALCHEMI studies, \citet{Harada2021} found that the cosmic ray ionization rate must be larger than \num{e3} $\zeta_0$ across the GMCs in the CMZ of \gal{} for chemical models to adequately explain the measured HCO$^+$ and HOC$^+$ emission. \citet{Holdship2021} also find that a cosmic ray ionization rate of $\zeta$ = \num{e3}-\num{e6} $\zeta_0$ is required to explain high column densities of C$_2$H. With \pwater and SO, we are now able to constrain the CRIR to within an order of magnitude for the first time, with all GMCs having a CRIR $\sim$\num{e4} $\zeta_0$ or \SI{e-13}{\per\second}.\par
It is useful to consider whether a consistent picture of the cosmic ray ionization rate in \gal{} is emerging from these studies. We therefore validate our inferred parameter distributions using previously reported abundances. To achieve this, we used the posterior samples provided by \texttt{Ultranest} to run UCLCHEM and probability distributions on the abundances of HCO$^+$, HOC$^+$, and C$_2$H according to our parameter distributions. One could include the line intensities of these species in the parameter inference but validating in this way allows us to effectively check our fits on held out data and evaluate the combination of \pwater and SO as a probe of the CRIR.\par
\begin{figure*}
    \centering
    \includegraphics[width=\textwidth]{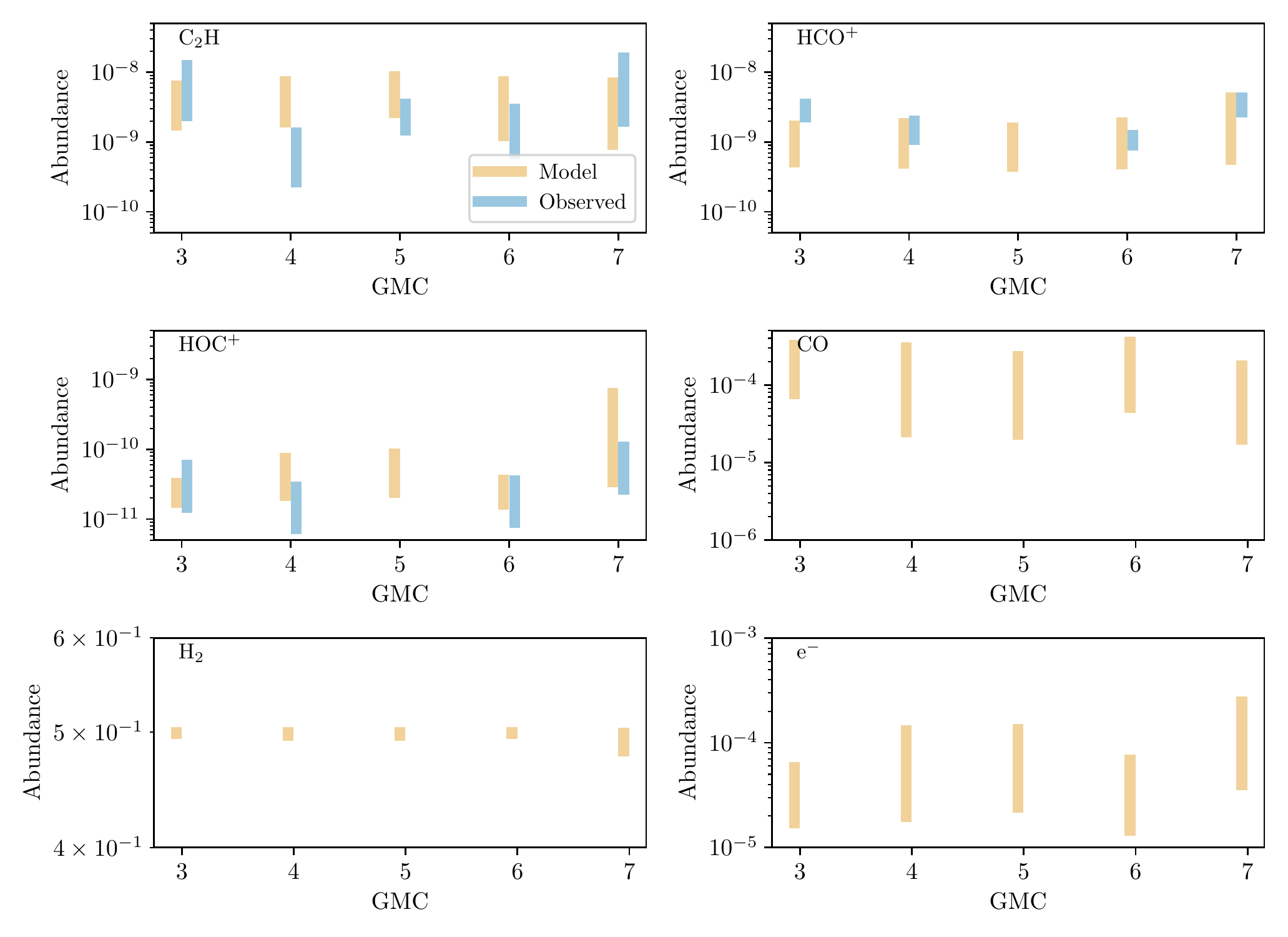}
    \caption{Abundance relative to total H nuclei of a selection of species for each region. The model range is the range of abundances obtained when sampling from the most likely portion of our parameter space using the constraints from this work. We also show observed ranges for some species which are the 1$\sigma$ ranges from \citet{Holdship2021} (C$_2$H) and \citealt{Harada2021} (HCO$^+$ and HOC$^+$). The CO, H$_2$ and e$^-$ abundances are shown to demonstrate that despite the high CRIR, the gas is largely molecular.}
    \label{fig:validation}
\end{figure*}

In analogy to a 1$\sigma$ interval, we define a likely range of model abundances of HCO$^+$, HOC$^+$, and C$_2$H by taking the 16th to 83rd percentile range of their respective distributions. We then compare these ranges to the observed abundances of those species as shown in Fig.~\ref{fig:validation}. In almost every case, the model and observed abundances overlap. The only exception is the C$_2$H abundance in GMC 4 but, statistically, one would not expect the 1$\sigma$ interval of every measurement to overlap with its true value so a small difference between the model and one measurement is not surprising. The overall agreement between the model and observations indicates that we have found a consistent picture of cosmic ray driven chemistry in the GMCs of NGC\,253.\par
As a further check, we also consider the abundance of H$_2$, CO and e$^-$  and include them in Fig.~\ref{fig:validation}. This allows us to check how strongly ionized the gas is. We find almost all H nuclei are in the form of H$_2$ indicating that the gas is still highly molecular under these conditions. Furthermore CO, we find that the CO abundance is similar to the elemental C abundance. This is important because \citet{Harada2021} reported a lower limit on the CRIR but limited it from above by the observation that a sufficiently high CRIR will dissociate CO, whilst observations of these regions indicate CO is abundant.\par 
\subsection{Alternatives to a High CRIR}
\citet{Holdship2021} found that either a high CRIR or a scenario where the GMCs were sufficiently clumpy to allow PDR chemistry over a large column density could reproduce the observed C$_2$H abundance. Similarly, \citet{Harada2021} found either PDR or cosmic ray dominated chemistry could reproduce observed abundances of HOC$^+$ and HCO$^+$. This ambiguity is somewhat resolved by the current study, as our \pwater and SO model fits greatly favour the high CRIR scenario, largely due to the fact \pwater{} is destroyed by UV but enhanced by cosmic rays.\par
In Sect.~\ref{sec:prelim}, we showed that the SO and H$_3$O$^+$ abundances are very low in PDR regions. However, when we use the posterior samples to generate model abundances for the likely parameter range, we find \pwater{} abundances between \num{e-10} and \num{e-9} which are too high for a PDR. Given that we have a consistent picture of CR-driven chemistry which reproduces the abundances of C$_2$H, \pwater, SO, HOC$^+$ and HCO$^+$, a UV-dominant scenario that works for some species but not others is disfavoured.\par
An alternative explanation for the high C$_2$H abundance was the presence of ubiquitous shocks in the GMCs. In Sect.~\ref{sec:prelim}, we found that in shock models, the \pwater{}:SO abundance ratio was approximately \num{e-4} across a wide range of physical parameters. Therefore, if our inferred ratio is much different to this, it is unlikely shocks dominate the chemistry of \pwater{} and SO. Using our posterior samples, we find that the ratio of these species' abundances varies between 0.2 and 22.9 across all regions when we consider the most likely parameter ranges given in Table~\ref{table:summary}. Since our inferred abundance ratio is always at least three orders of magnitude larger than would be found in a shock, it seems unlikely that it is in fact shock chemistry controlling the abundance of these species. \par
One final possibility that has not yet been discussed in this work is that of X-ray driven chemistry. Due to their similar ionizing effect and weak attenuation with column density, X-rays can have very similar effects to cosmic rays on the chemistry of a gas \citep{Viti2014}. Therefore, it is possible some or all of the chemistry ascribed to cosmic rays in this work is due to X-rays.\par
Whilst no X-ray flux was included in the PDR models discussed in Sect.~\ref{sec:prelim}, UCL\_PDR does treat X-rays if a flux is provided. We therefore run a model of GMC 5 using our best fit gas density and an estimate of the largest X-ray flux that can be motivated from observations of X-ray sources around the GMCs. \citet{Lehmer2013} observed an X-ray source which they call Source B to be in the brightest in \gal{} with a luminosity of $\sim$\SI{e39}{\erg}. It is also very close to GMC 5, at a distance of $\sim$\SI{20}{\parsec}. Combining these values, we obtain a flux of \SI{0.05}{\erg\per\centi\metre\squared\per\second}. By modelling a cloud that is subjected to this flux, we can observe the maximum effect of X-rays on our GMC chemistry.\par
\begin{figure}
    \centering
    \includegraphics[width=0.5\textwidth]{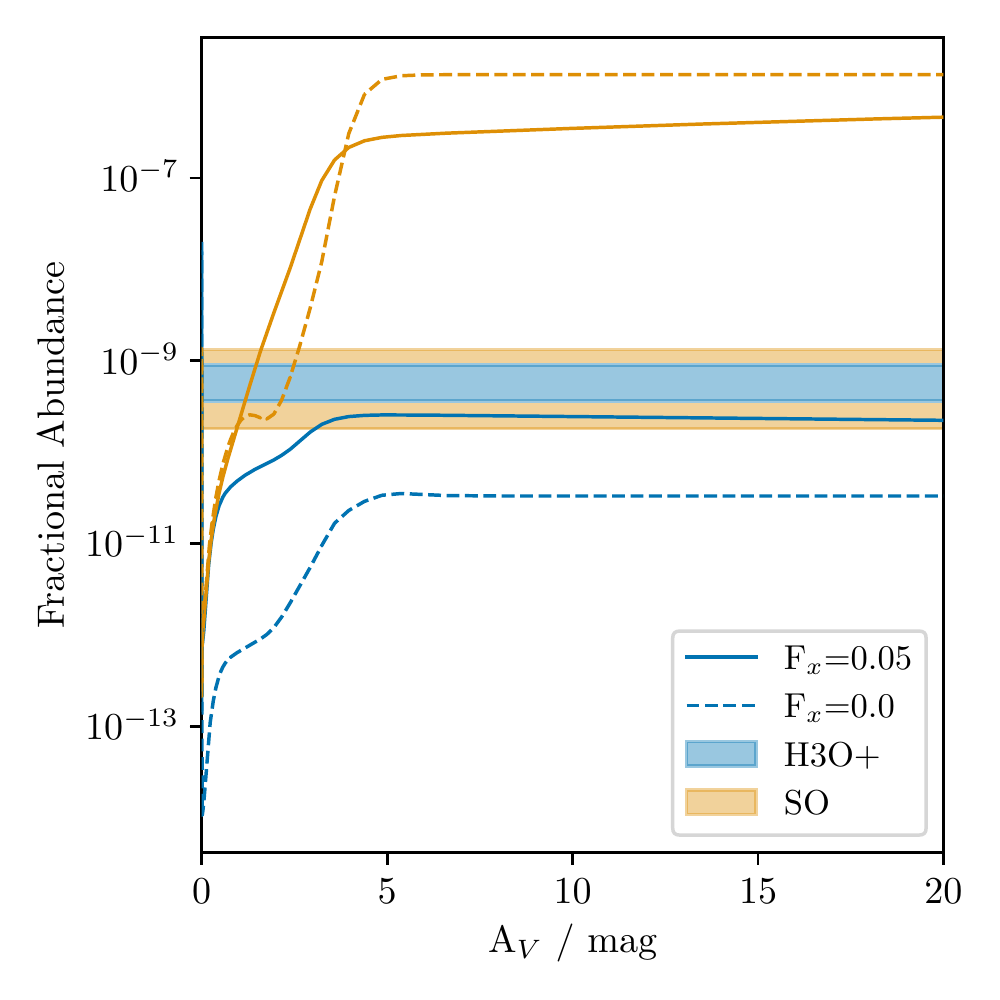}
    \caption{Lines show the modelled fractional abundance of SO and \pwater as a function of visual extinction under an external X-ray flux of \SI{0.05}{\erg\per\centi\metre\squared\per\second} and under no X-ray flux. The shaded regions show the measured 67\% most likely range of fractional abundances.}
    \label{fig:xray}
\end{figure}
We plot the results in Fig.~\ref{fig:xray}. We find that increasing the X-ray flux increases the \pwater{} abundance and decreases the SO abundance in a similar manner to cosmic rays. However, the effect of an X-ray flux of this magnitude is not sufficient to bring the model abundances in line with observations. One could argue the \pwater{} abundance is reasonable given the uncertainties in chemical models but the SO abundance is two orders of magnitude higher than that found from our RADEX fits. Since rectifying this with an XDR model would require an input X-ray flux that is orders of magnitude higher than could be expected from any observed X-ray source, we conclude that cosmic rays must be driving most SO destruction.\par
\section{Conclusions}
\label{sec:conclusion}
Previous work has shown the cosmic ray ionization rate (CRIR) in the CMZ of \gal{} is large but has failed to constrain it to within an order of magnitude. Therefore, in this work emission from SO and \pwater\ toward several positions in the CMZ of \gal{} was analysed due to the fact that their abundances are strongly tied to the CRIR.\par
We find that we can constrain the CRIR in these regions and that in every location it is $\sim$\SI{e-13}{\per\second} (\num{e4}$\zeta_0$), and is extremely unlikely to fall outside the range \SIrange[]{e-14}{e-12}{\per\second}. We validate this result by using our inferred parameter values to model the abundances of previously detected species that have been used as probes of the CRIR and find that we reproduce the observed abundances.\par
Previous ALCHEMI studies have been unable to rule out alternative processes that could explain the observed abundances of their CRIR probes. However, our inferred chemical abundances strongly disfavour shock dominated chemistry and the high \pwater\ abundance found in the GMCs of \gal{} disfavours PDR chemistry. Further, the low SO abundance is unattainable with purely X-ray driven chemistry. Since all of the CRIR tracers presented by ALCHEMI can be consistently modelled through cosmic ray dominated chemistry, we now consider this to be the most likely scenario.\par 
We find no evidence for variation in the CRIR between the GMCs we have modelled. Whilst GMCs 3 and 7 have the lowest most likely CRIR values, the uncertainties in those values are sufficiently large that they overlap considerably with the CRIR values derived for other GMCs. Given that the model already fails to capture the complexity of the data, it is likely a more accurate model is required rather than more data in order to better constrain the CRIR in each GMC.\par
\acknowledgments
We thank F. Priestley for helpful discussions on XDR modelling. We also thank the anonymous reviewer for their insightful comments on this manuscript. This work is part of a project that has received funding from the European Research Council (ERC) under the European Union’s Horizon 2020 research and innovation programme MOPPEX 833460. This paper makes use of the following ALMA data: ADS/JAO.ALMA\#2017.1.00161.L, ADS/JAO.ALMA\#2018.1.00162.S., and 2016.1.01285.S. ALMA is a partnership of ESO (representing its member states), NSF (USA) and NINS (Japan), together with NRC (Canada), MOST and ASIAA (Taiwan), and KASI (Republic of Korea), in cooperation with the Republic of Chile. The Joint ALMA Observatory is operated by ESO, AUI/NRAO and NAOJ.  The National Radio Astronomy Observatory is a facility of the National Science Foundation operated under cooperative agreement by Associated Universities, Inc. The work of YN was supported by NAOJ ALMA Scientific Research Grant No.~2017-06B and JSPS KAKENHI grant No.~JP18K13577.
V.M.R. and L.C. have received funding from the Comunidad de Madrid through the Atracci\'on de Talento Investigador (Doctores con experiencia) Grant (COOL: Cosmic Origins Of Life; 2019-T1/TIC-15379).  L.C. has also received partial support from the Spanish State Research Agency (AEI; project number PID2019-105552RB-C41).
\par
\bibliography{references.bib}{}
\bibliographystyle{aasjournal}



\appendix

\section{The ALCHEMI \pwater and SO Data}
\label{sec:AppendixIntFigs}
\begin{figure*}[hbpt]
\centering
\includegraphics[trim=0mm 0mm 0mm 0mm, clip, scale=0.61]{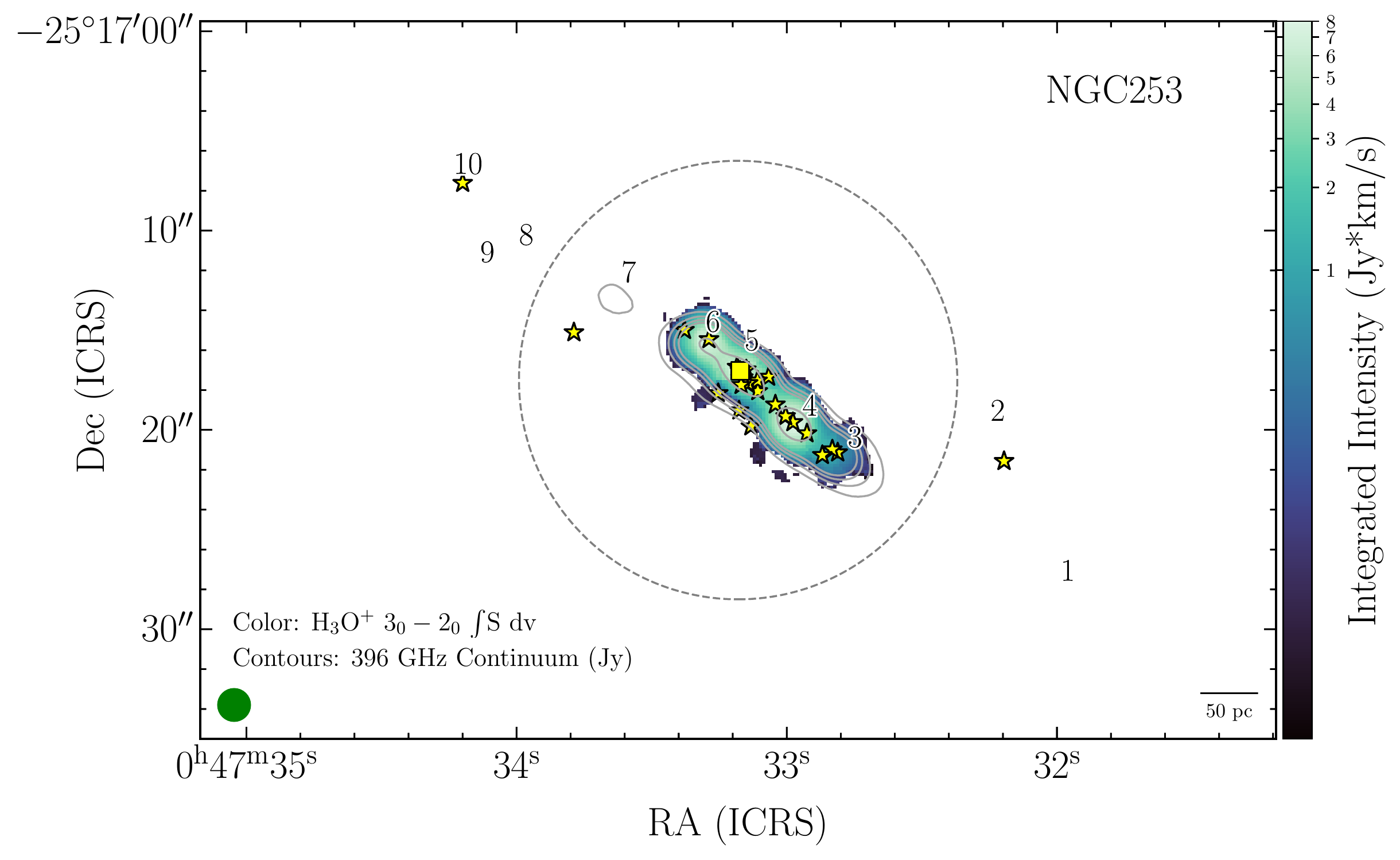}\\
\includegraphics[trim=0mm 0mm 0mm 0mm, clip, scale=0.61]{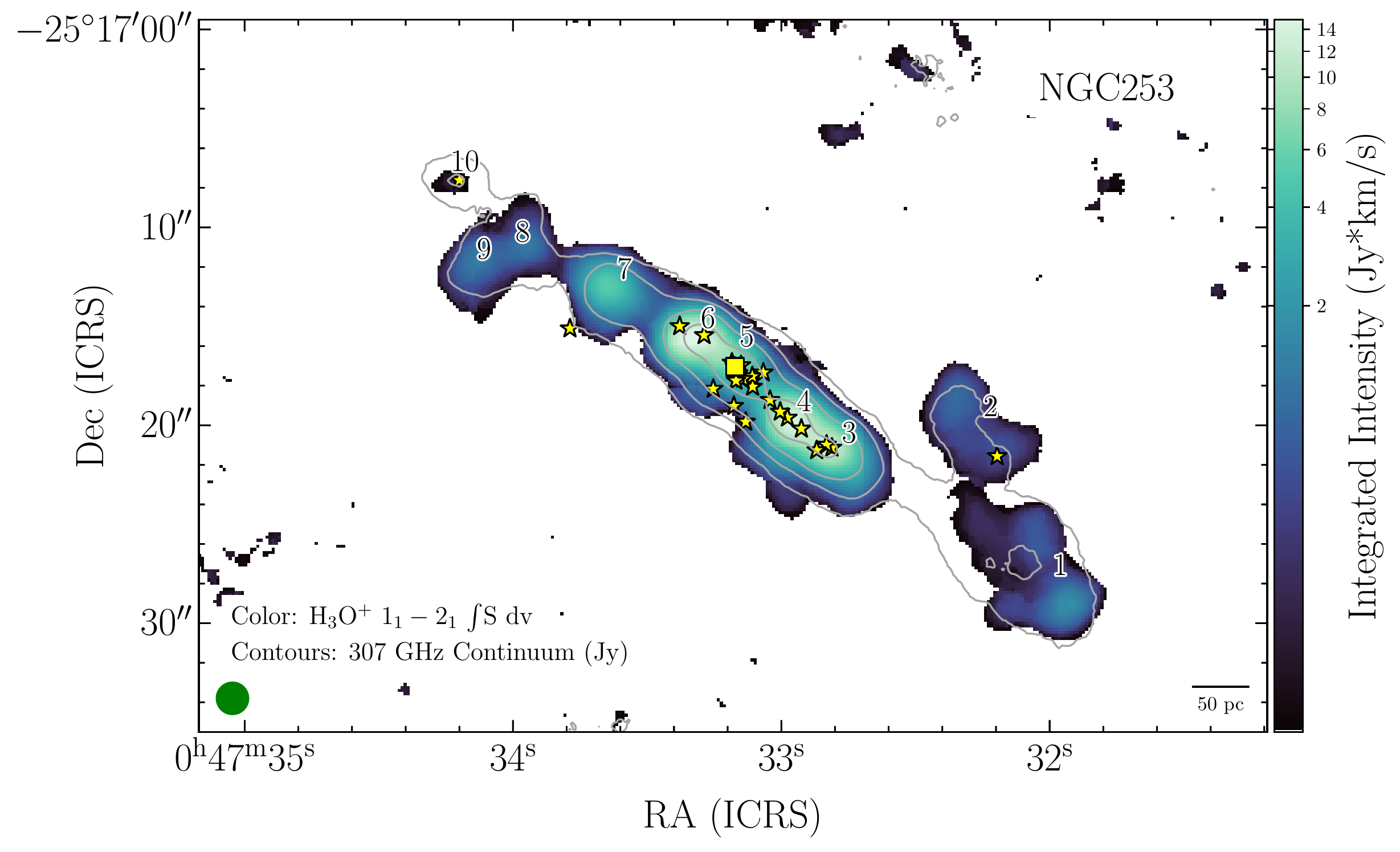}
\caption{Additional \pwater{} integrated intensity (moment 0) images toward NGC\,253.  Markings, intensity scaling, and contours in each panel same as for Figure~\ref{fig:H3Op320221IntSO3423Int}.  The continuum RMS values for the transitions shown are 10.0 (top) and 1.0 (bottom) mJy/beam, respectively.  Note that the field of view for the H$_3$O$^+$ $3_0-2_0$ transition is 22 arcsec centered at $\alpha=00^h47^m33.134^s$,  $\delta= -25^{\circ}17'19.68''$ (ICRS, shown as a grey dashed circle; see Section~\ref{sec:pwater3020}).}
\label{fig:H3OpIntAdd}
\end{figure*}

\begin{figure*}[hbpt]
\centering
\includegraphics[trim=0mm 10mm 0mm 0mm, clip, scale=0.33]{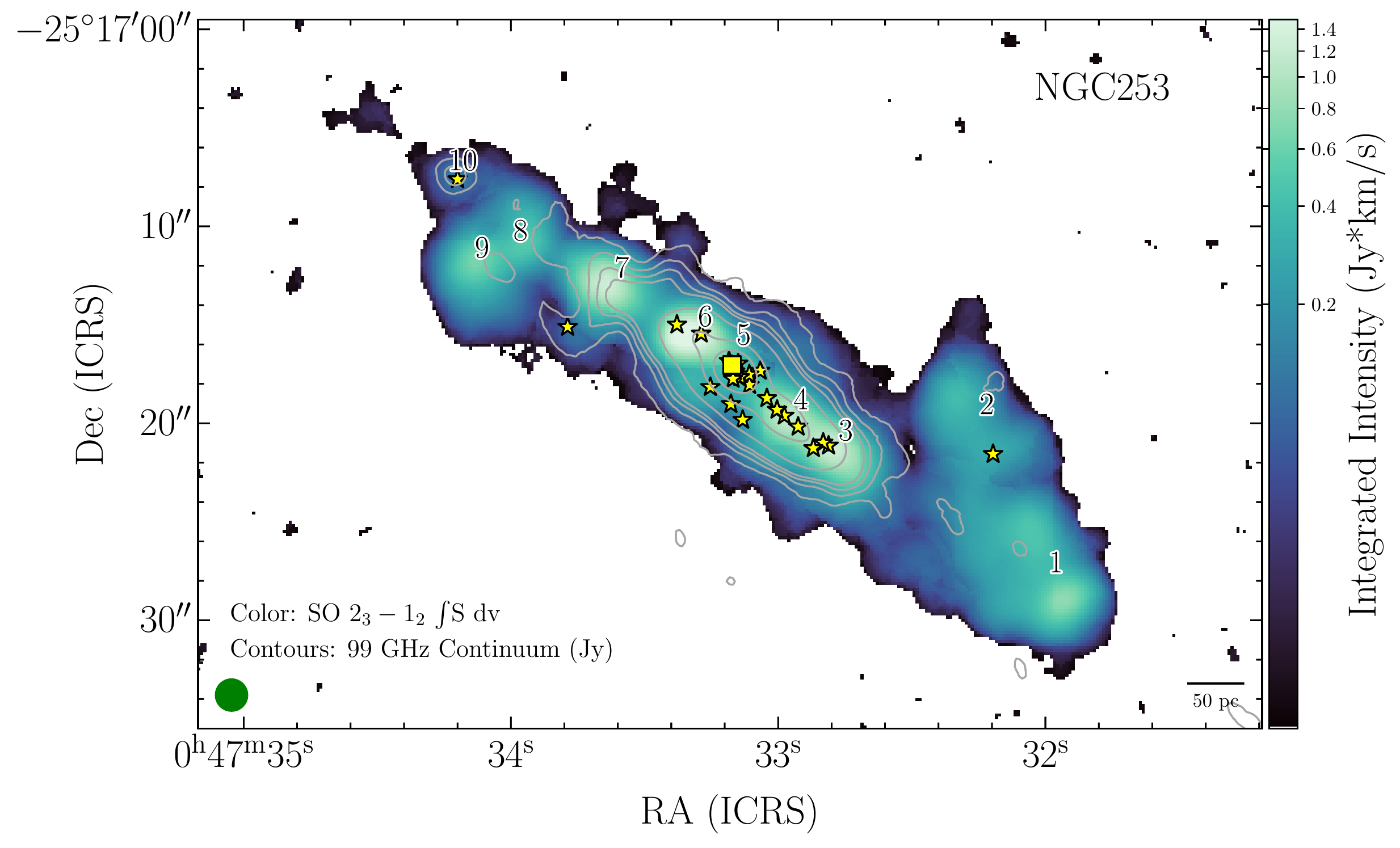}
\includegraphics[trim=10mm 10mm 0mm 0mm, clip, scale=0.33]{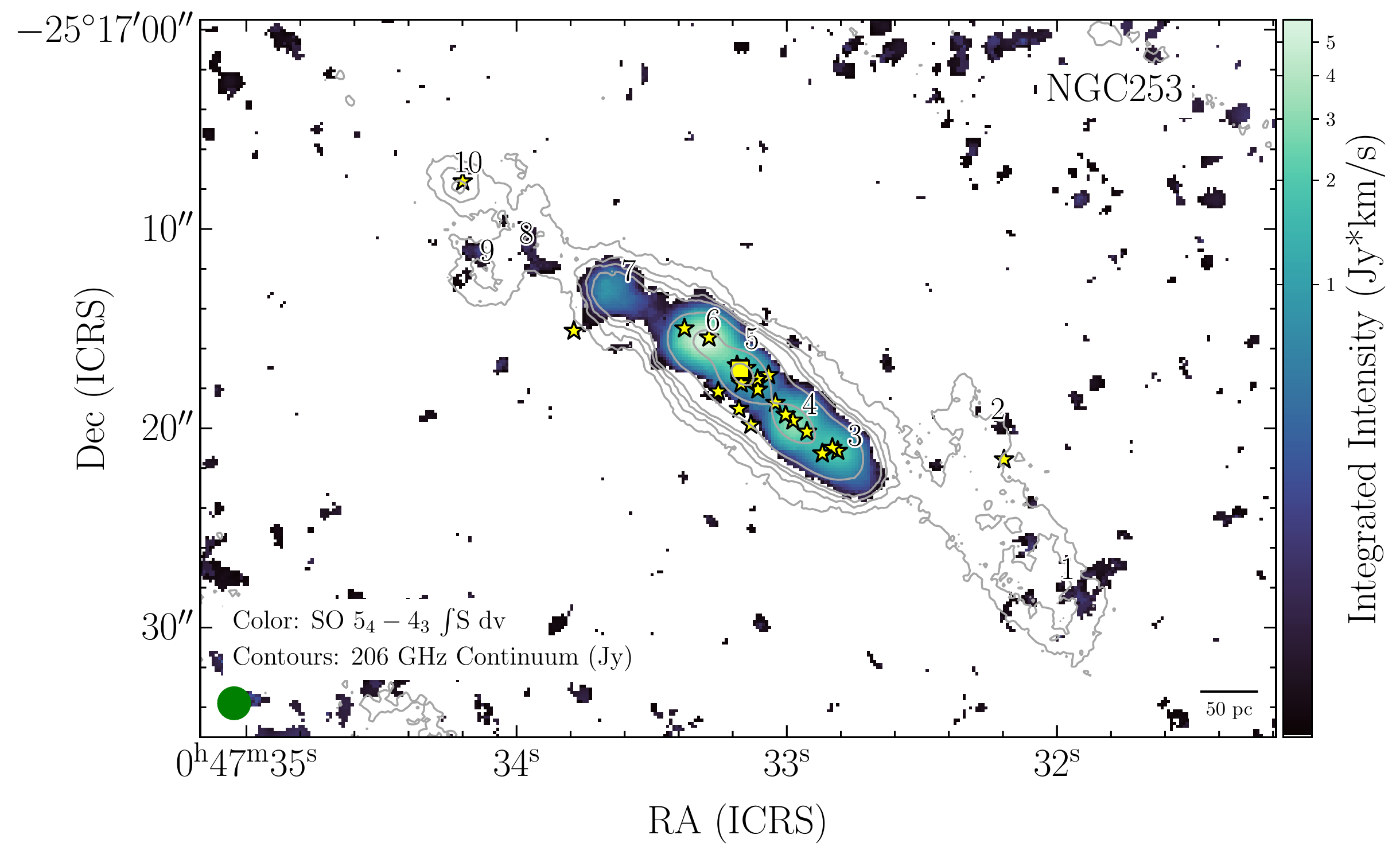}\\
\includegraphics[trim=0mm 0mm 0mm 0mm, clip, scale=0.33]{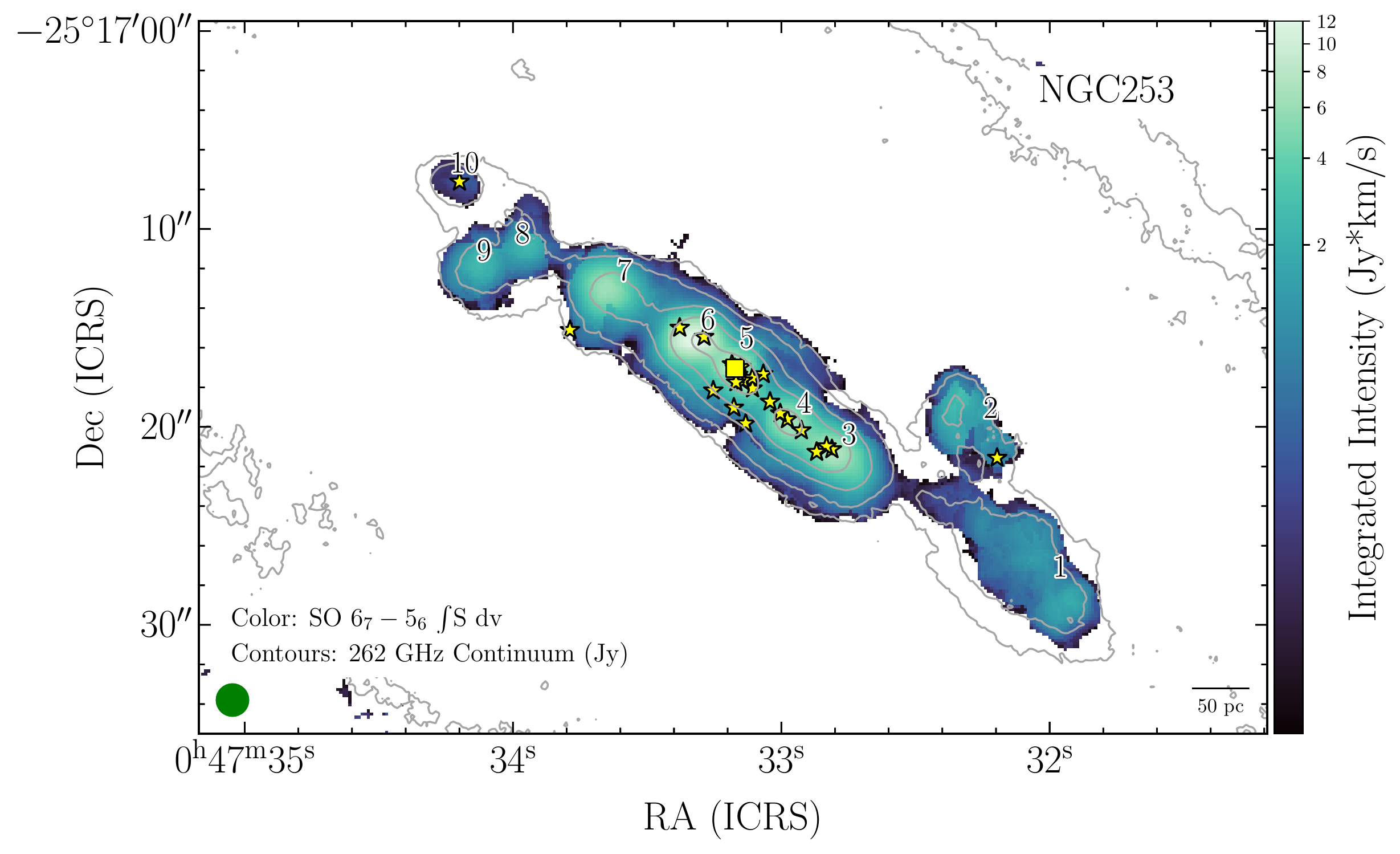}
\includegraphics[trim=10mm 0mm 0mm 0mm, clip, scale=0.33]{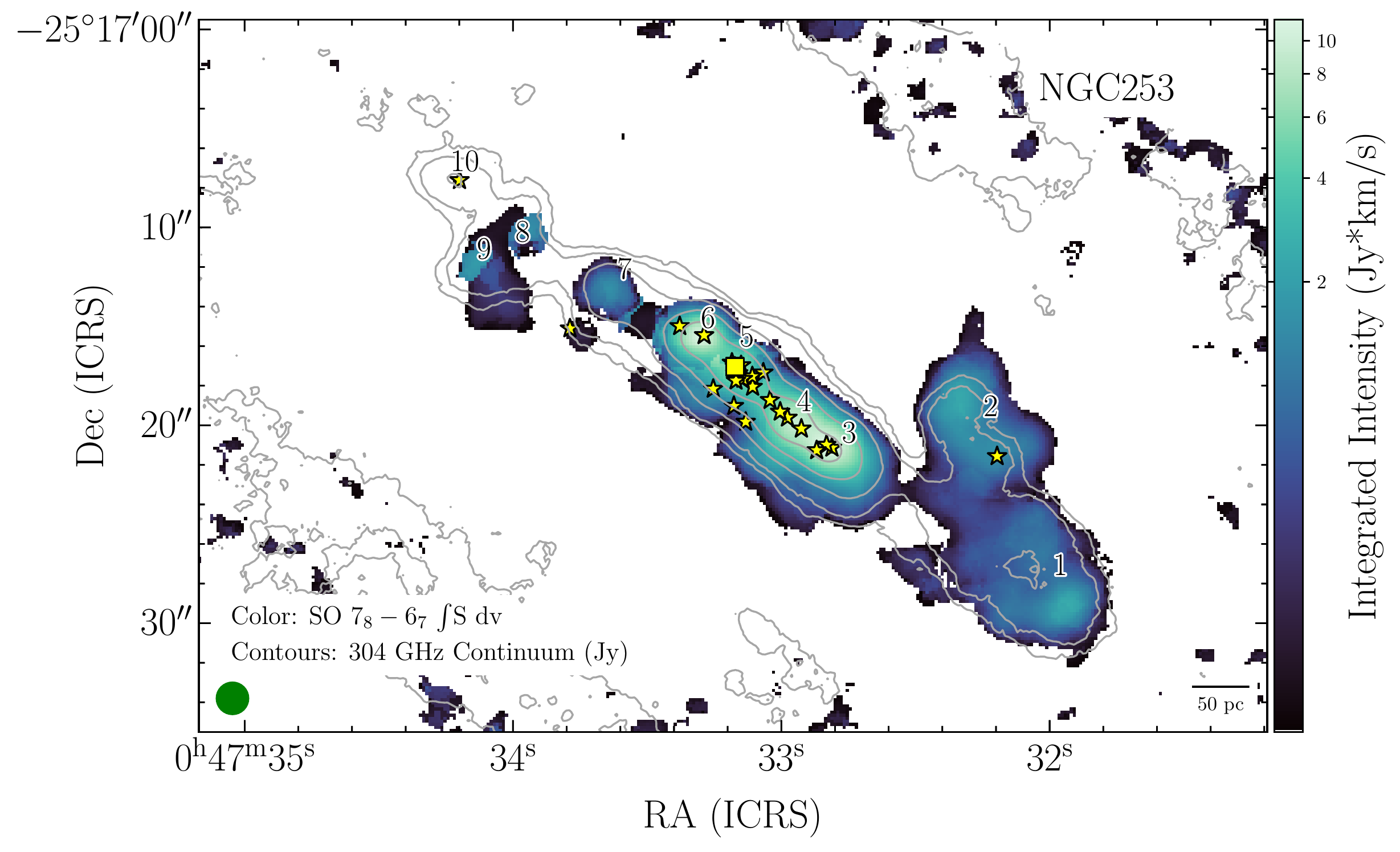}
\caption{Additional SO integrated intensity (moment 0) images toward NGC\,253.  Markings, intensity scaling, and contours in each panel same as for Figure~\ref{fig:H3Op320221IntSO3423Int}.  The continuum RMS values for the transitions shown are 0.17 (top left), 0.33 (top right), 0.33 (lower left), and 0.33 (lower right) mJy/beam, respectively.}
\label{fig:SOIntAdd}
\end{figure*}

In this section, we present the data analysed in this work. In Figures~\ref{fig:H3OpIntAdd} and ~\ref{fig:SOIntAdd}, we show additional moment 0 maps of NGC 253's CMZ. We also provide an example of the data included in the supplementary data table. Table~\ref{table:data-example} shows the transitions and measured line intensities detected in GMC 4, the equivalent values for the other GMCs are available in the online article.

\begin{deluxetable*}{rllrrrrrrr}
\tablewidth{0pt}
\tablecaption{Detected transitions, their properties and measured intensities for GMC 4.}
\label{table:data-example}
\tablehead{
    \colhead{Region}&
    \colhead{Species\tablenotemark{a}}&
    \colhead{Transition\tablenotemark{a}}&
    \colhead{Frequency}&
    \colhead{E$_U$}&
    \colhead{RA}&
    \colhead{Dec}&
    \colhead{Corrected Integrated}&
    \colhead{Integrated Intensity}&
    \colhead{Interloper Correction}
    \\
    \colhead{}&
    \colhead{}&
    \colhead{}&
    \colhead{(GHz)}&
    \colhead{(K)}&
    \colhead{}&
    \colhead{}&
    \colhead{Intensity (K km/s)}&
    \colhead{Uncertainty (K km/s)}&
    \colhead{}
}
\startdata
  4 &  H3Op & 1.1.1-2.1.0 & 307.192410 & 79.5 & 11.8874 & -25.28888 &  13.509450 & 8.817727 & 0.23 \\
  4 &  H3Op & 3.2.0-2.2.1 & 364.797430 &  139.3 & 11.8874 & -25.28888 &  61.744377 &  10.782524 & 0.86 \\
  4 &  H3Op & 3.0.1-2.0.0 & 396.272410 &  169.1 & 11.8874 & -25.28888 &  18.992481 & 3.812451 & 1.00 \\
  4 &  SO & 2.3-1.2 &  99.299870 &  9.2 & 11.8874 & -25.28888 &  47.964352 & 7.209908 & 1.00 \\
  4 &  SO & 1.1-0.1 & 286.340152 & 15.2 & 11.8874 & -25.28888 & 0.591833 & 0.354076 & 1.00 \\
  4 &  SO & 3.4-2.3 & 138.178600 & 15.9 & 11.8874 & -25.28888 &  45.781056 & 6.875517 & 1.00 \\
  4 &  SO & 2.2-1.1 &  86.093950 & 19.3 & 11.8874 & -25.28888 &  13.188165 & 2.028540 & 1.00 \\
  4 &  SO & 2.2-1.2 & 309.502444 & 19.3 & 11.8874 & -25.28888 & 0.973434 & 0.366215 & 1.00 \\
  4 &  SO & 3.2-2.1 & 109.252220 & 21.1 & 11.8874 & -25.28888 &  20.209262 & 3.069187 & 1.00 \\
  4 &  SO & 4.5-3.4 & 178.605403 & 24.4 & 11.8874 & -25.28888 &  32.354178 & 6.111951 & 1.00 \\
  4 &  SO & 3.3-2.2 & 129.138923 & 25.5 & 11.8874 & -25.28888 &  20.025210 & 3.028742 & 1.00 \\
  4 &  SO & 4.3-3.2 & 158.971811 & 28.7 & 11.8874 & -25.28888 &  28.146298 & 4.251155 & 1.00 \\
  4 &  SO & 4.4-3.3 & 172.181403 & 33.8 & 11.8874 & -25.28888 &  20.281011 & 3.075113 & 1.00 \\
  4 &  SO & 5.6-4.5 & 219.949442 & 35.0 & 11.8874 & -25.28888 &  34.456452 & 5.173552 & 1.00 \\
  4 &  SO & 5.4-4.3 & 206.176005 & 38.6 & 11.8874 & -25.28888 &  27.974328 & 4.237034 & 1.00 \\
  4 &  SO & 5.4-4.4 & 100.029640 & 38.6 & 11.8874 & -25.28888 & 4.935441 & 0.872624 & 1.00 \\
  4 &  SO & 5.5-4.4 & 215.220653 & 44.1 & 11.8874 & -25.28888 &  18.644894 & 2.810657 & 1.00 \\
  4 &  SO & 6.7-5.6 & 261.843721 & 47.6 & 11.8874 & -25.28888 &  28.611328 & 5.660309 & 0.76 \\
  4 &  SO & 6.5-5.4 & 251.825770 & 50.7 & 11.8874 & -25.28888 &  20.277435 & 5.747689 & 0.53 \\
  4 &  SO & 6.5-5.5 & 136.634799 & 50.7 & 11.8874 & -25.28888 &  26.897403 & 4.049473 & 1.00 \\
  4 &  SO & 6.6-5.5 & 258.255826 & 56.5 & 11.8874 & -25.28888 &  18.888161 & 2.867306 & 1.00 \\
  4 &  SO & 7.8-6.7 & 304.077844 & 62.1 & 11.8874 & -25.28888 &  38.849257 & 5.843079 & 1.00 \\
  4 &  SO & 7.6-6.5 & 296.550064 & 64.9 & 11.8874 & -25.28888 &  17.696586 & 2.673156 & 1.00 \\
  4 &  SO & 7.7-6.6 & 301.286124 & 71.0 & 11.8874 & -25.28888 &  17.585570 & 2.662084 & 1.00 \\
  4 &  SO & 8.9-7.8 & 346.528481 & 78.8 & 11.8874 & -25.28888 &  18.149714 & 2.795038 & 1.00 \\
  4 &  SO & 8.7-7.6 & 340.714155 & 81.2 & 11.8874 & -25.28888 &  14.417925 & 2.531354 & 0.87 \\
  4 &  SO & 8.8-7.7 & 344.310612 & 87.5 & 11.8874 & -25.28888 &  11.763875 & 1.843587 & 0.99 \\
  4 &  SO & 9.8-8.8 & 254.573628 & 99.7 & 11.8874 & -25.28888 &  29.755258 & 4.477287 & 1.00 \\
  \enddata
\tablenotetext{a}{For machine readable format "$^+$" replaced by "p" and transitions listed as J.K$^{-1}$.K$^{+1}$ and J.K for H$_3$O$^+$ and SO, respectively.}
\end{deluxetable*}
\section{Corner Plots}
\label{sec:corners}
In this section, we include the corner plots for GMCs 3, 5, 6, and 7 which are not in the main article.
\begin{figure*}[hbpt]
    \centering
    \includegraphics[width=\textwidth]{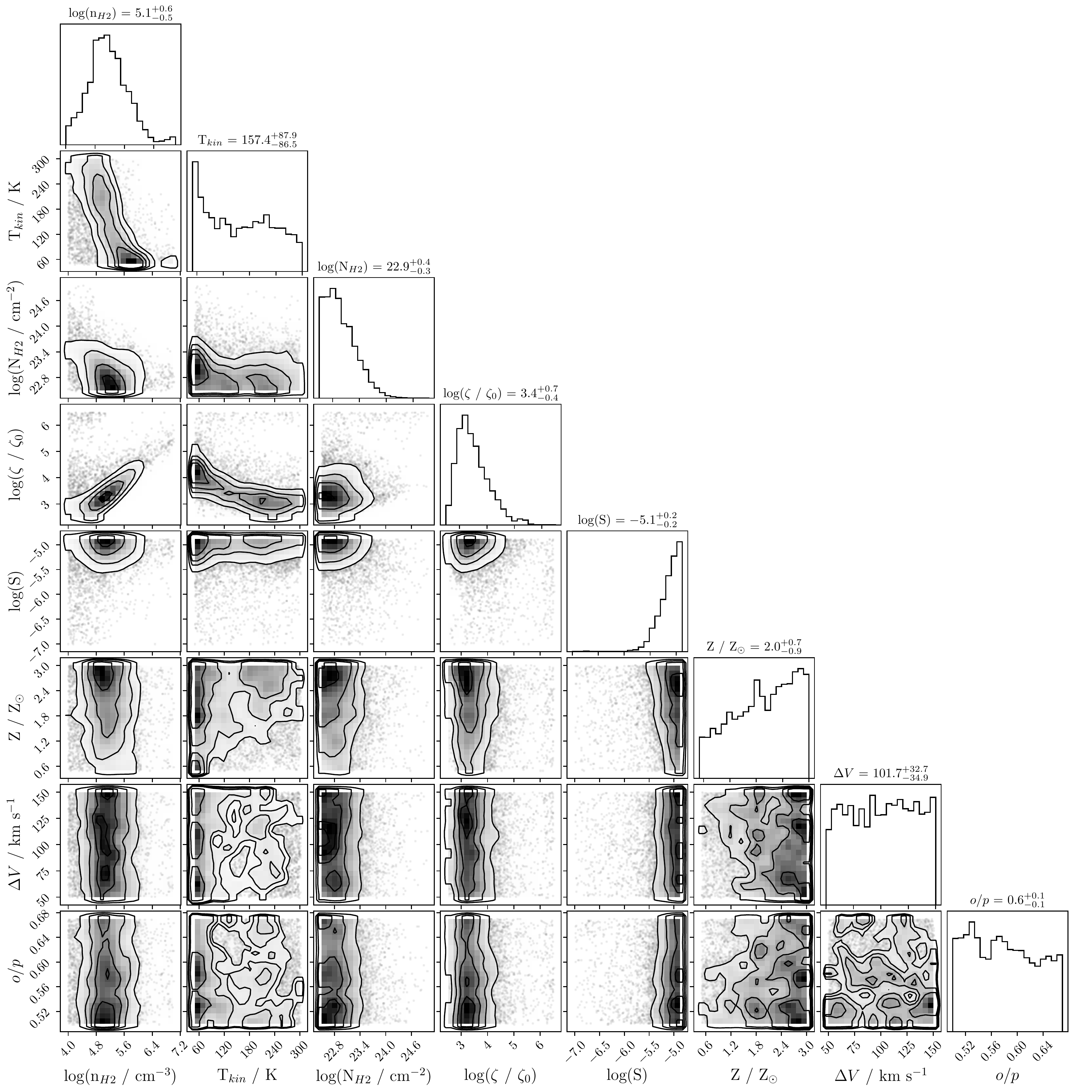}
    \caption{Corner plot showing the marginalized posterior probability distribution of every parameter and the joint distributions of all parameter pairs for GMC 3.}
    \label{fig:corner3}
\end{figure*}
\begin{figure*}
    \centering
    \includegraphics[width=\textwidth]{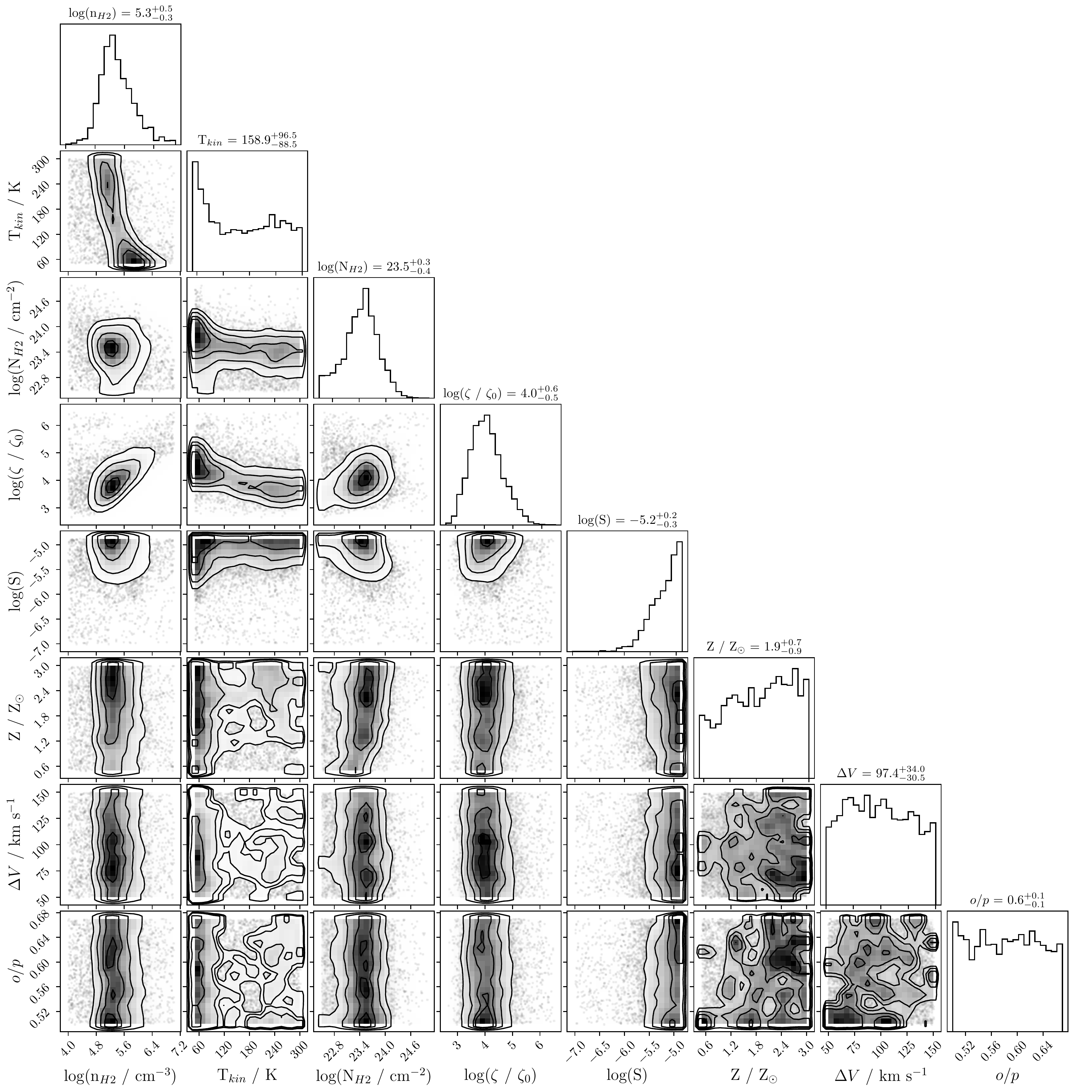}
    \caption{Similar to Fig.~\ref{fig:corner3} for GMC 5}
    \label{fig:corner5}
\end{figure*}
\begin{figure*}
    \centering
    \includegraphics[width=\textwidth]{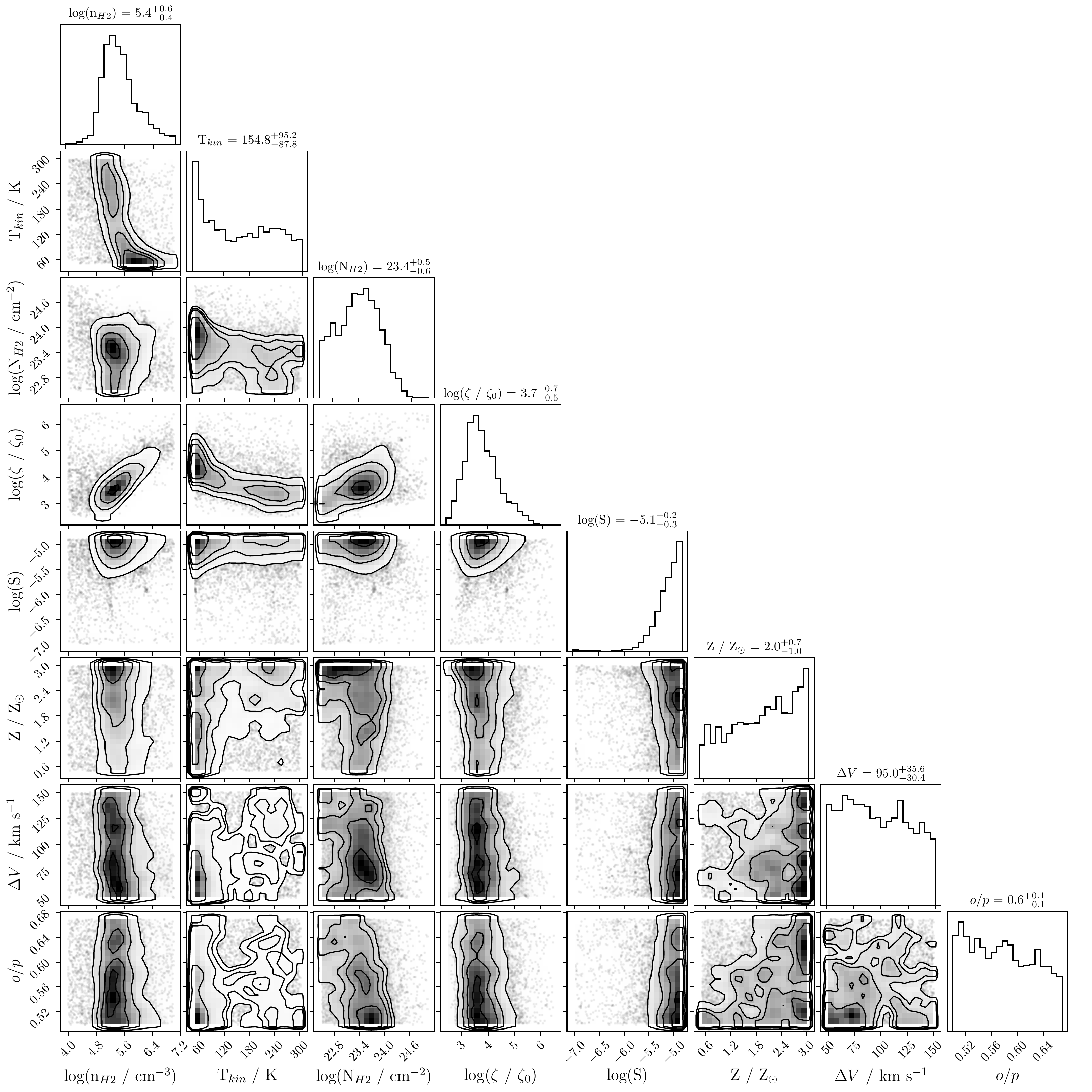}
    \caption{Similar to Fig.~\ref{fig:corner3} for GMC 6}
    \label{fig:corner6}
\end{figure*}
\begin{figure*}
    \centering
    \includegraphics[width=\textwidth]{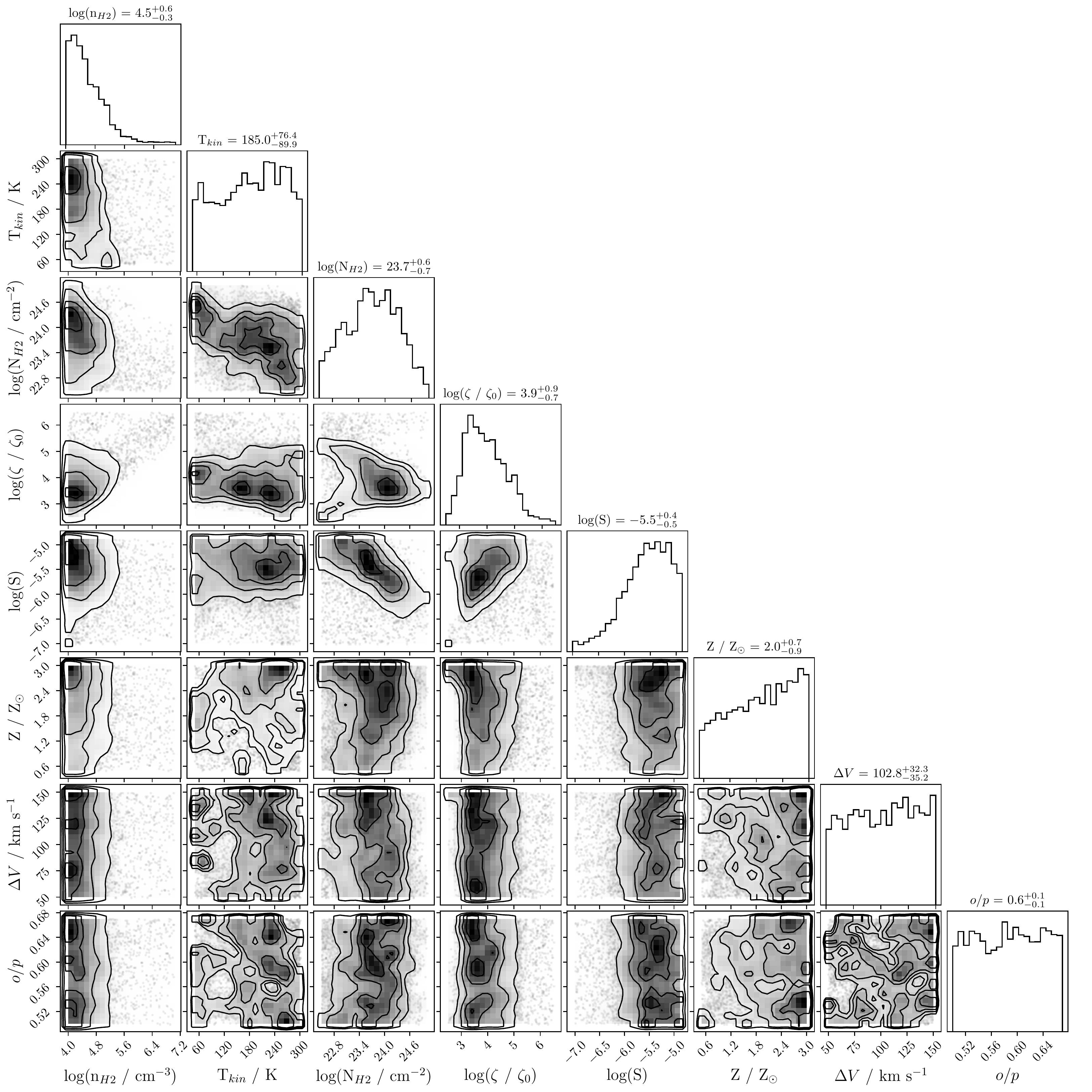}
    \caption{Similar to Fig.~\ref{fig:corner3} for GMC 7}
    \label{fig:corner7}
\end{figure*}

\end{CJK*}
\end{document}